\documentclass[a4paper,fleqn,usenatbib]{mnras} 

\usepackage{newtxtext,newtxmath}

\usepackage[T1]{fontenc}
\usepackage{ae,aecompl}

\usepackage{graphicx}	
\usepackage{amsmath}	
\usepackage{amssymb}	
\usepackage{graphics}
\usepackage{xcolor}
\usepackage{textcomp}     

\usepackage{natbib}
\usepackage{pdflscape}
\usepackage[export]{adjustbox}
\usepackage[percent]{overpic}

\usepackage{hyperref}
\usepackage{rotating}
\usepackage{dblfloatfix}    
\usepackage{multicol}

\usepackage{pdflscape}
\usepackage{afterpage}

\usepackage[final]{changes}
\makeatletter
\@namedef{Changes@AuthorColor}{black}
\colorlet{Changes@Color}{black}
\makeatother

\newcommand{\Ha}{H$\alpha$ }

\newcommand{\dg}{$\Delta${HI}MS}  
\newcommand{\dgg}{$\Delta${\H2}MS}

\newcommand{\hi}{H{\sc i}}

\newcommand\FramedBoxx[2]{%
  \setlength\fboxsep{0pt}
  \setlength{\fboxsep}{0pt}%
  \setlength{\fboxrule}{0pt}%
  \fbox{\parbox[t][0pt][c]{#1}{\raggedleft\scriptsize #2}}}

\def\msun{M$_\odot$}

\def\mhi{M$_\textrm{\hi}$}
\def\mH2{M$_\textrm{\H2}$}

\def\H2{H$_2$}
\def\L100{L$_{100}$}
\def\Lo160{L$_{160}$}
\def\L250{L$_{250}$}
\def\L350{L$_{350}$}
\def\L500{L$_{500}$}

\def\Mst{M$_*$}

\title[xGASS: Cold gas in the Transition Zone]{xGASS: Cold gas content and quenching in galaxies below
  the star forming main sequence}

\author[Janowiecki et al.]{ 
Steven~Janowiecki,$^{1,2}$\thanks{E-mail: steven.janowiecki@icrar.org (SJ)}
Barbara~Catinella,$^{1,3}$
Luca~Cortese,$^{1,3}$
Amelie Saintonge,$^{4}$
\newauthor
\, Jing Wang$^{5}$
\\
\scriptsize
$^{1}$International Center for Radio Astronomy Research (ICRAR), M468,
The University of Western Australia, 35 Stirling Highway,  Crawley,
WA, 6009, Australia\\
\scriptsize
$^{2}$University of Texas, Hobby-Eberly Telescope, McDonald Observatory, TX 79734, USA\\
\scriptsize
$^{3}$ARC Centre of Excellence for All Sky Astrophysics in 3
Dimensions (ASTRO 3D)\\
\scriptsize
$^{4}$Department of Physics and Astronomy, University College London, Gower Street, London WC1E 6BT, UK\\
\scriptsize
$^{5}$Kavli Institute for Astronomy and Astrophysics, Peking University, Beijing 100871, China
}

\date{ Accepted 2020 January 13. Received 2020 January 3; in original form 2019 April 12}

\pubyear{2020}

\begin{document}
\label{firstpage}
\pagerange{\pageref{firstpage}--\pageref{lastpage}}
\maketitle

\begin{abstract}

We use \hi \, and \H2 global gas measurements of galaxies from xGASS
and xCOLD~GASS to investigate quenching paths of galaxies below the
star formation main sequence (SFMS). We show that the population of
galaxies below the SFMS is not a 1:1 match with the population of
galaxies below the \hi \, and \H2 gas fraction scaling
relations. Some galaxies in the transition zone (TZ) 1-sigma below the
SFMS can be 
as \hi-rich as those in the SFMS, and have on average
longer gas depletion timescales. We find evidence for environmental
quenching of satellites, but central galaxies in the TZ defy simple
quenching pathways. Some of these so-called ``quenched'' galaxies may
still have significant gas reservoirs and be  unlikely to deplete them
any time
soon. As such, a correct model of galaxy quenching cannot be inferred
with SFR (or other optical observables) alone, but must include
observations of the cold gas. We also find that internal
structure (particularly, the spatial distribution of old and
young stellar populations) plays a significant role in regulating the
star formation of gas-rich isolated TZ galaxies, suggesting the
importance of bulges in their evolution. 


\end{abstract}




\section{Introduction}

Observations have shown a strong correlation between global estimates
of star formation rate (SFR) and stellar mass (\Mst) for populations
of galaxies, commonly referred to as the ``star forming main
sequence'' \citep[SFMS,][]{brinchmann04, elbaz07, salim07}. The
tightness of the SFMS suggests that star formation proceeds in a
fairly universal and relatively calm mode, without significant
deviations in SFR \citep{noeske07}. In an idealized completely
isolated galaxy, secular evolution will produce this smooth and
gradual evolution as stellar mass grows, gas is consumed, star
formation diminishes, and optical color becomes redder. Here, the SFR
would depend largely on the available gas, and the two would decline
hand-in-hand until the galaxy becomes red and fully dead.

In reality, galaxy evolution is more complex, as most galaxies do not
spend their entire lives in isolation. They can merge or interact with
other galaxies or dark matter halos, which open up further
evolutionary pathways and can have dramatic effects on their gas
content and star formation. However, the tightness of the SFMS across
cosmic time \citep[e.g.,][]{daddi07,magdis10,whitaker12} puts a strong
constraint on the maximum amplitude of deviations a star-forming
galaxy can 
experience. In particular, the shape and redshift evolution of the
SFMS are consistent with a picture where star formation is regulated
by cold gas availability, star formation efficiency
(SFE=SFR/M$_\textrm{gas}$), and feedback \citep{bouche10, lilly13,
  saintonge16}.

As galaxies descend from the SFMS towards the red sequence (RS)
they cross through a transitional zone (TZ), historically referred to
as the ``green valley'' because of their intermediate optical and
ultra-violet (UV) colors  \citep{martin07, wyder07,
  salim07}. However, this transitional population of galaxies is
better described by its intermediate specific star formation rates
(sSFR=SFR/\Mst), rather than the original mass- and dust-dependent
color selection criteria \citep{cortese12,woo13,salim14}.

Most of the quenching studies discussed above have relied on UV,
optical, near-infrared (NIR), and spectroscopic observations of stellar
populations to characterize the evolutionary pathways between the SFMS
and the red sequence. However, observations
of the 
neutral atomic (\hi) and molecular (\H2) cold gas reservoirs are
required to fully 
understand the evolutionary potential of galaxies below the SFMS. Even
if its stellar populations appear
``quenched,'' a galaxy may possess a significant gas reservoir and
still have significant potential for further evolution or growth.

Previous studies of cold gas in the TZ are relatively
few in number, as it becomes difficult to detect in galaxies below the
SFMS. \citet{cortese09} found that TZ galaxies are \hi-deficient in
the high-density environment of the Virgo Cluster and \hi-rich in
lower density environments. Independent of
environment, \citet{schiminovich10} found that the average \hi \, gas
depletion time (\mhi/SFR) was nearly constant on and below the SFMS
(although with a large scatter),
suggesting a common regulator of both gas supply and star
formation. More recently, \citet{saintonge16} found a fairly constant
  SFE (inverse of a depletion time) for massive galaxies in the SFMS,
  and argued that their overall cold gas supply drives their location
  in the SFR-\Mst \, plot, and that their evolution is not driven by
  bottlenecks in the conversion from atomic 
  to molecular gas. 

In this work we quantify the cold gas properties of galaxies
in the TZ below the SFMS to characterize their evolutionary
pathways. This effort is made possible with extensive \hi \,
observations from the recently completed 
low-mass e\textbf{x}tensions of the 
\textit{GALEX} Arecibo SDSS Survey \citep[xGASS,][]{catinella18} and
its molecular gas counterpart, CO Legacy Database for GASS
\citep[xCOLD~GASS,][]{saintonge17}. Combined, these surveys include
measurements of the cold atomic and molecular 
gas content in a representative, stellar mass-selected sample
of galaxies, including a significant 
number from the gas-poor regime. By including the cold gas reservoirs
in galaxies across the SFR-\Mst \, plane, we
consider the potential for future star formation in galaxies that
might currently appear ``quenched'' in terms of their star
formation. While these TZ galaxies appear to be in the
gloaming of their lives, many have significant reservoirs of cold gas
that can support sustained or increased future star formation, and
have not been truly quenched yet.

This paper is organized as follows: In Section~\ref{sec:MS} we
quantify the SFMS and gas fraction scaling relations of the xGASS
sample. Section~\ref{sec:off} quantifies the gas properties in
galaxies departing from the SFMS with extreme depletion
times. Section~\ref{sec:gv} characterizes the population of galaxies
in the TZ below the SFMS, and Section~\ref{sec:discussion} discusses
these results. Finally, we summarize our main results and discuss
future work in Section~\ref{sec:summary}.

\begin{figure*}
\centering
\includegraphics[width=0.97\textwidth]{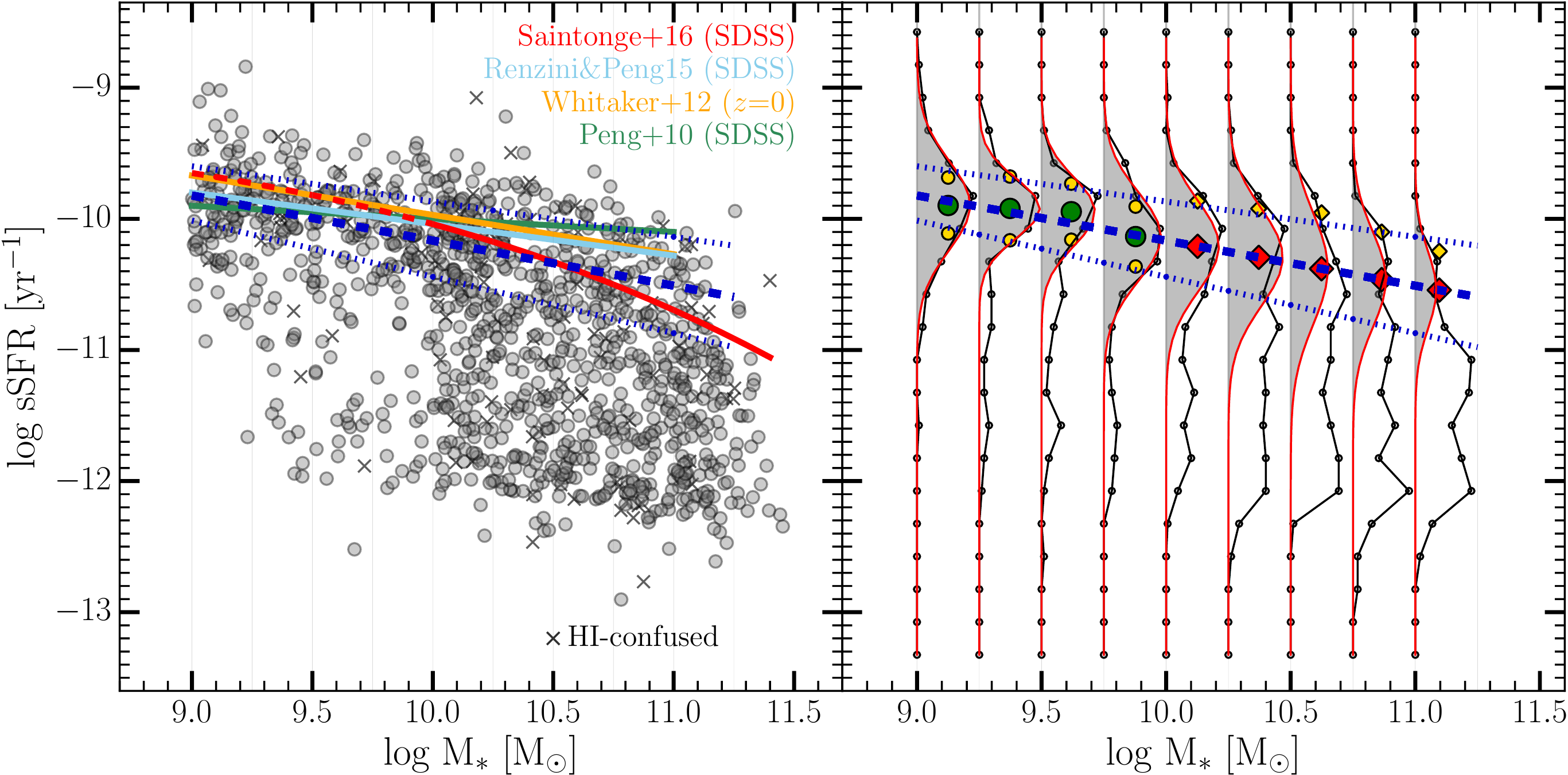}
\caption{Left panel: sSFR is plotted against stellar mass for our
  sample (grey circles), including \hi-confused galaxies (grey
  crosses). Blue thick dashed and dotted lines show our SFMS relation and
  its 1$\sigma$ scatter, and solid colored lines show SFMS relations from
  other groups.
  Right panel: a visual summary of our SFMS fitting method. Within
  each stellar mass bin, black lines and points show the sSFR
  distributions. Grey histograms show the best-fitting Gaussians in
  each bin. The four green dots show the modes of the unconstrained
  Gaussian fits at low masses, and the red diamonds show the
  extrapolation to higher stellar mass. The yellow symbols show the
  best-fitting 1-sigma widths in each bin. Our best-fitting SFMS
  relation ($\pm$1$\sigma$) is shown as a blue dashed (dotted) line(s).
    \label{fig:sfms}}
\end{figure*}

\section{\texorpdfstring{\MakeLowercase{x}}xGASS}
\label{sec:MS}

We use galaxies from the xGASS sample \citep{catinella18}, which is the
extension to lower stellar masses of the sample from
\citet{catinella10,catinella13}. xGASS includes $\sim$1200 galaxies in the local
Universe ($0.01$\textless$z$\textless$0.05$), evenly sampling the
stellar mass interval $10^9$\textless\Mst/\msun\textless$10^{11.5}$
with no other selection criteria. The galaxies have been observed in
21cm by Arecibo until \hi \, is detected or until an upper limit of a
few percent is reached on the \hi \, gas fraction
(\mhi/\Mst). Practically, the gas fraction limit varies as a function
of stellar mass \citep[see Section~2.1.1 of ][]{catinella18}, and
galaxies with \hi \, detections below this gas fraction limit are
considered non-detections in this work. This survey represents the
most sensitive \hi \, observations of a local representative galaxy
sample to date.

In addition to the neutral atomic 21cm observations, CO(1-0)
observations exist for $\sim$40\% of xGASS galaxies 
and are
included in xCOLD~GASS \citep{saintonge17}, which is the extension to
lower stellar masses of the sample from \citet{saintonge11}. Combining
these molecular observations with the \hi \, data provides a complete
inventory of the cold gas fuel for star formation throughout the xGASS
sample.

Beyond these observations of cold gas, the xGASS sample has a
rich set of complementary data. SDSS optical photometry come from
Data Release 7 \citep[DR7,][]{dr7}, and stellar masses come from the
value-added
catalog\footnote{\href{http://www.mpa-garching.mpg.de/SDSS/DR7/}{http://www.mpa-garching.mpg.de/SDSS/DR7/};
  we used the
  improved stellar masses from\\
  \href{http://home.strw.leidenuniv.nl/~jarle/SDSS/}{http://home.strw.leidenuniv.nl/$\sim$jarle/SDSS/}} 
provided by the Max Planck Institute for Astrophysics (MPA) and Johns
Hopkins University (JHU), which assume a \citet{chabrier03} initial
mass function. Further multi-wavelength observations include images in
ultra-violet (UV) from 
the \textit{Galaxy Evolution Explorer}
\citep[\textit{GALEX},][]{martin05,morrissey07} and in 
mid-infrared (MIR) from the \textit{Wide-field Infrared Survey Explorer}
\citep[\textit{WISE},][]{wright10}. As described in
\citet{janowiecki17}, these UV+MIR observations are used to derive
total UV+IR SFRs for xGASS galaxies, fully accounting for the direct
emission of the young stellar populations in UV, and for the
re-processed dust emission in MIR. 

In this work we  use these UV+IR SFRs to define the xGASS SFMS
and determine the position of each galaxy above or below this
relation. Analogously, we  use the
\hi \, and \H2 scaling relations from \citet{catinella18} and
\citet{saintonge17} 
to quantify how far above or below the typical gas scaling relations
galaxies are at a given stellar mass. The details of our SFMS
determination and the gas fraction scaling relations we use are
described in the following subsections.

\subsection{Fitting the SFMS}

In order to explore galaxies that depart from the SFMS, we must first
define it in our sample. The xGASS SFMS was first presented in
\citet{catinella18} (Equation~2); we discuss its derivation in
more detail in this work. 
The left panel of Figure~\ref{fig:sfms} shows
sSFR as a function of stellar mass for the xGASS sample. We include 
$\sim$100~galaxies that have unreliable \hi \, measurements due to
possible source confusion, although they will be excluded from any
subsequent analysis that requires \hi \, masses \citep[see Appendix~A
  in][for further discussion of source confusion]{catinella18}. 
We are interested in characterizing the SFMS relation as well as its
width, to determine how far a galaxy has departed from it. In 
the following paragraphs, we describe our extrapolated half-mirrored
Gaussian method to determine the xGASS SFMS and its width.

First, we divide the sample into 0.25~dex-wide stellar mass bins
starting from 
\Mst=10$^9$\msun, as separated by the thin grey vertical lines in the
left panel of Figure~\ref{fig:sfms}. In each of the four lowest bins
(9~$\le$~log~\Mst/\msun~$\le$~10), the sample is dominated
by star-forming galaxies, so we fit unconstrained Gaussians to the sSFR
distributions. Using a linear 
extrapolation of the modes from these four low mass bins, we fit Gaussians to
the higher mass bins with a fixed center and allow only their widths to
vary. In this higher mass bin we fit a 1-sided Gaussian distribution
to only the galaxies that lie above the SFMS to avoid being affected
by the increasingly dominant red
sequence. After this Gaussian fitting is complete, we fit the
$\pm$1$\sigma$ values with a simple linear relation to parametrize the
width of the SFMS as a function of stellar mass. Our
best-fitting SFMS and its width ($\sigma$, in dex) are given in the form:
\begin{equation}
\log \frac{\textrm{SFR}_\textrm{MS}}{\textrm{\Mst}} [\textrm{yr}^{-1}] =
m_\text{SFMS} \left( \log \frac{\textrm{\Mst}}{\textrm{\msun}} - 9 \right)
 +b_\text{SFMS}
\end{equation}
\begin{equation}
\sigma_\textrm{MS} [\textrm{dex}] = 
m_\text{SFS} \left( \log \frac{\textrm{\Mst}}{\textrm{\msun}} - 9 \right)
+ b_\text{SFS}.
\end{equation}
\noindent
where
$m_\textrm{SFMS}$=$-0.344\pm0.101$,
$b_\textrm{SFMS}$=$-9.822\pm0.057$,
$m_\textrm{SFS}$=$0.088\pm0.028$,
and $b_\textrm{SFS}$=$0.188\pm0.036$. 

Other determinations of the SFMS are also shown in the left panel of
Figure~\ref{fig:sfms} and are broadly consistent with ours, although
some divergence is seen at higher masses where the SFMS becomes less
populated. Each
SFMS comes from a different sample of galaxies and uses a different SFR
indicator (or indicators), different definitions of
star-forming galaxies, and different fitting
methods. While all formulations are similar, we adopt the xGASS SFMS
for consistency with our previous work. 

We define distance above/below the SFMS as $\Delta$SFMS:
\begin{equation}
\Delta\textrm{SFMS} = \log\frac{\textrm{SFR}}{\textrm{\Mst}}
                 [\text{yr}^{-1}] - 
\log\frac{\textrm{SFR}_\textrm{MS}}{\textrm{\Mst}} [\text{yr}^{-1}].
\end{equation}
\noindent
Note that this difference is essentially a (logarithmic) ratio of two
sSFRs and as such is dimensionless. We determine $\Delta$SFMS in
dimensionless logarithmic (dex) units without regard for the changing
width of the SFMS as a 
function of \Mst. While the SFMS is narrowest ($\sim$0.19~dex) at our
lowest stellar masses 
and gets broader with increasing mass, the change is relatively
small. Further, using dex units makes it easier to interpret
$\Delta$SFMS values at different stellar masses in a physically
meaningful way.

\begin{figure*}
\centering
\includegraphics[width=0.90\textwidth]{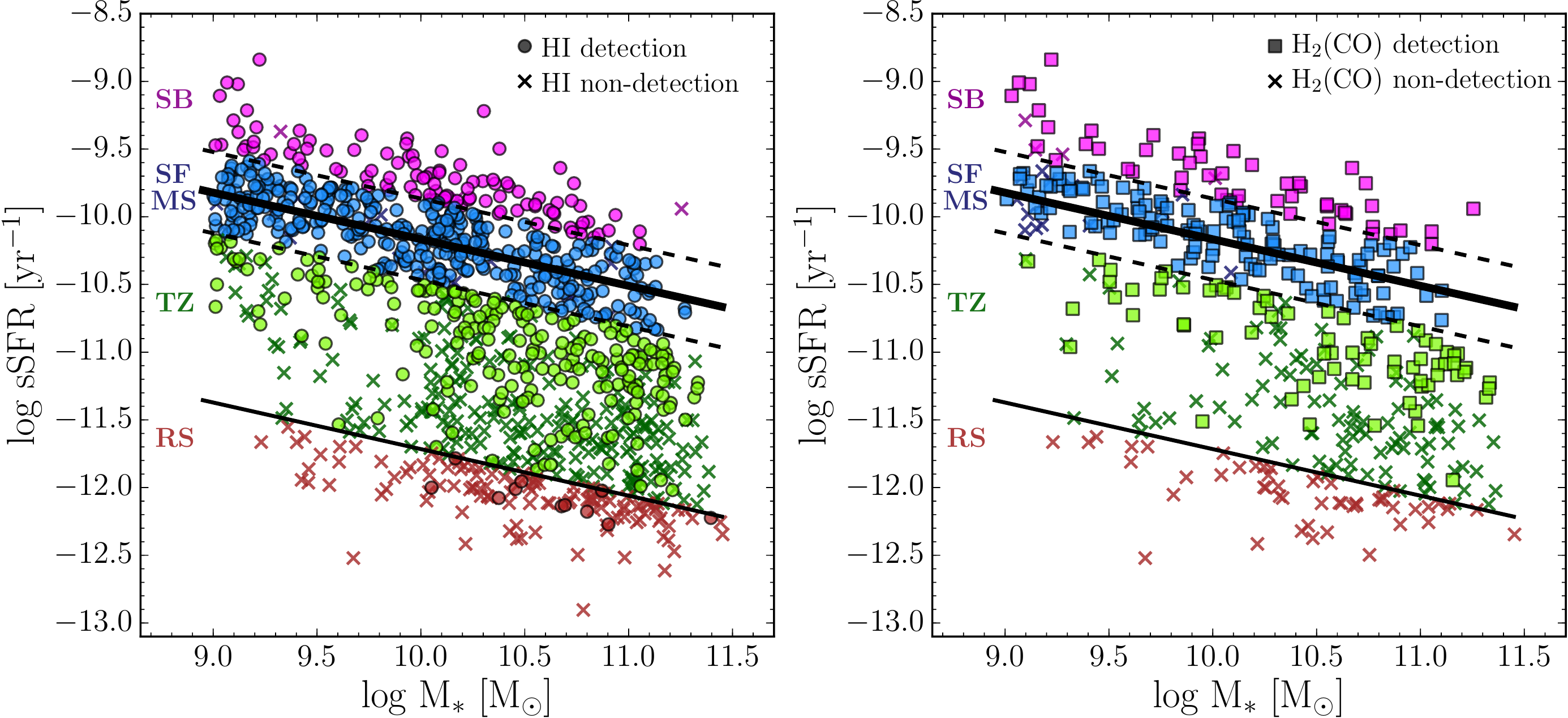}\\
\hspace{5pt}
\includegraphics[width=0.89\textwidth]{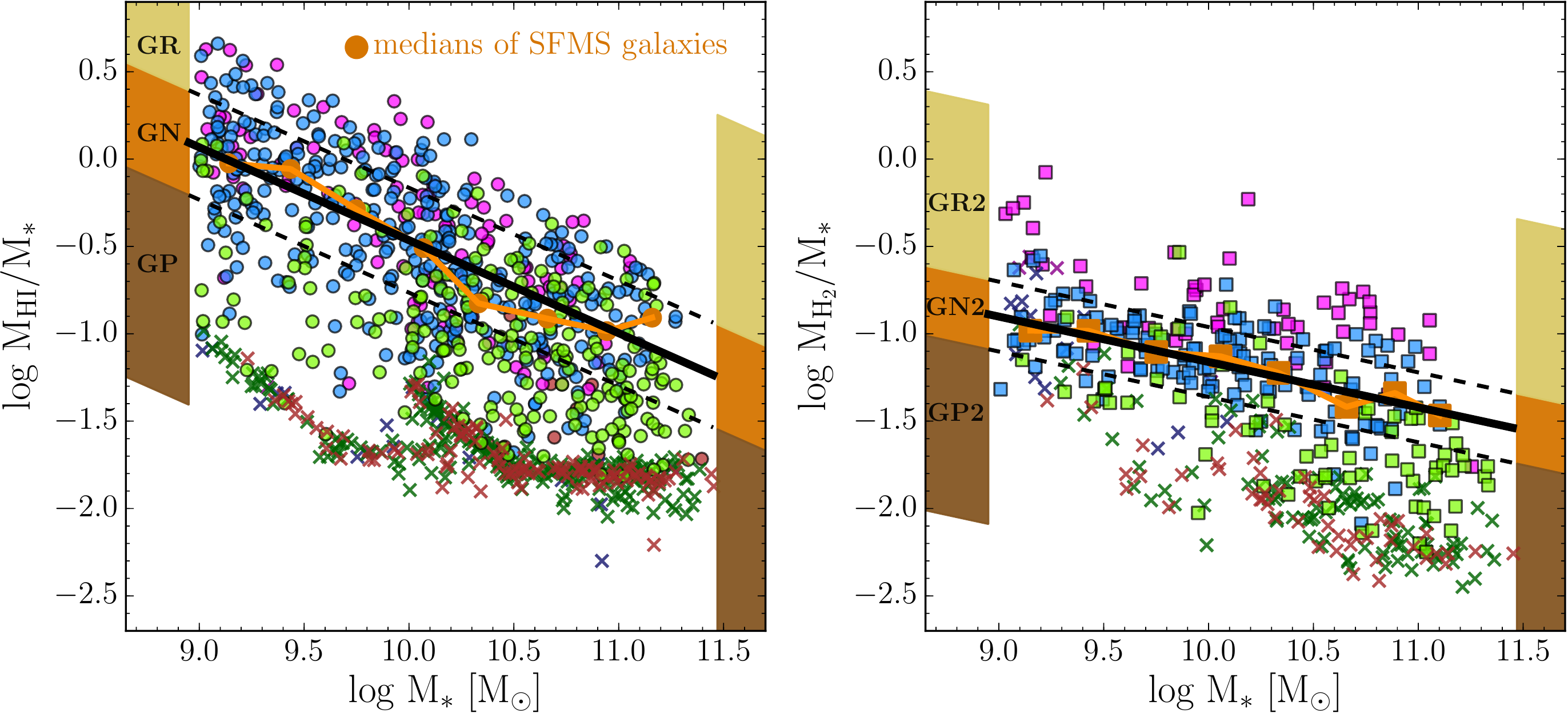}
  \caption{Top two panels: sSFR vs.~\Mst \, is color coded by
    position 
    above/on/below the SFMS, with shapes based on their \hi \,
    (left panel, circles) or CO (right panel, squares) detection
    status, where crosses indicate  non-detections.  
    Galaxies with potentially confused \hi \, observations are not included.
    Black lines show the xGASS SFMS with $\pm$0.3~dex dashed lines and
    the threshold of the RS.
  Bottom two panels: \hi \, and \H2 gas fractions use the same
  colors and shapes as the top panels. The large orange
  points show median \hi \, and \H2 
  gas fractions of galaxies from the SFMS in bins of stellar mass, 
  black lines show linear fits to those averages, and dashed lines
  show $\pm$0.3~dex and $\pm$0.2~dex for \hi \, and \H2 gas fractions,
  respectively. 
    \label{fig:sfgf}}
\end{figure*}

\subsection{Cold gas fraction relations of SFMS galaxies}

Next we apply a similar approach to characterize the typical \hi \,
and \H2 gas fraction scaling relations for the galaxies in the SFMS we
have just defined. 
We carry out a linear fit to the
median gas fraction in each bin of stellar mass for galaxies in the
SFMS. These binned averages (orange dots) and best-fit relations
(black lines) are 
shown in the bottom two panels of Figure~\ref{fig:sfgf}, and are
parameterized in the same form as the SFMS, as:
\begin{equation}
\log \frac{\textrm{\mhi}_\textrm{,MS}}{\textrm{\Mst}}  =
m_\text{GFMS} \left( \log \frac{\textrm{\Mst}}{\textrm{\msun}} - 9 \right)
 - b_\text{GFMS}
\end{equation}
\begin{equation}
\log \frac{\textrm{\mH2}_\textrm{,MS}}{\textrm{\Mst}}  =
m_\text{GF2MS} \left( \log \frac{\textrm{\Mst}}{\textrm{\msun}} - 9 \right)
 - b_\text{GF2MS}.
\end{equation}
\noindent
where  
$m_\textrm{GFMS}$=$-0.53\pm0.06$,
$b_\textrm{GFMS}$=$0.07\pm0.01$,
$m_\textrm{GF2MS}$=$-0.26\pm0.03$, and
$b_\textrm{GF2MS}$=$-0.90\pm0.18$.

As with the SFMS, we define the distance above/below these \hi \, and
\H2 scaling relations as \dg \, and \dgg,
respectively. As with $\Delta$SFMS, this is a logarithmic ratio in
units of dex, as shown below:
\begin{equation}
$\dg$ = 
\log\frac{\textrm{\mhi}}{\textrm{\Mst}} - 
\log\frac{\textrm{\mhi}_\textrm{,MS}}{\textrm{\Mst}}
\end{equation}
\begin{equation}
$\dgg$
= \log\frac{\textrm{\mH2}}{\textrm{\Mst}} - 
\log\frac{\textrm{\mH2}_\textrm{,MS}}{\textrm{\Mst}}.
\end{equation}

Since galaxies in the SFMS do not correspond to a similarly
tight sequence in either the \hi \, or \H2 gas fraction relations, we
do not fit Gaussians to determine a width. Instead, we adopt widths of
0.3~dex (0.2~dex) 
to separate galaxy populations that have \mhi \, (\mH2) values which
are similar to the expectation from the simple scaling relation; these
widths correspond to the 1$\sigma$ standard deviations of the SFMS
population which is detected in each panel. 
The bottom two panels of Figure~\ref{fig:sfgf}
show these regions of parameter space. 
Note that in this work we consistently treat the \hi \, and
molecular gas masses separately. The gas fraction relations of the
total (\hi \, + molecular) gas would be qualitatively similar to the
\hi \, relations as the gas reservoirs of the SFMS and TZ galaxies are
dominated 
by the \hi \, component. For a detailed analysis and
discussion of scaling relations based on a combined total cold gas
mass, see Section~4.2 of \citet{catinella18}.

\begin{figure*}
\centering
{\Large \hspace{1.2cm} \hi \hspace{5.8cm} \H2~(CO)}
\includegraphics[width=0.735\textwidth]{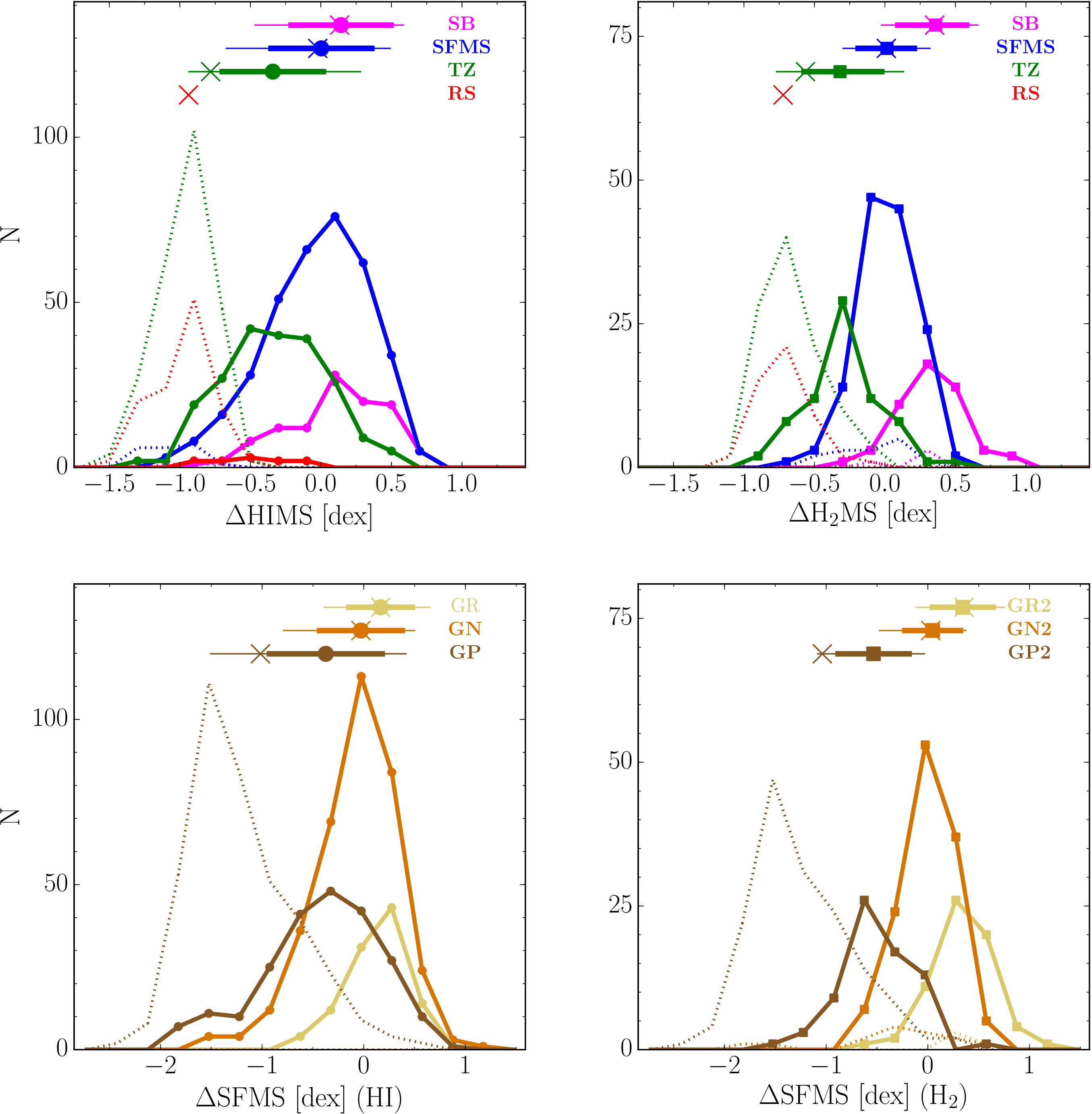}
  \caption{Top panels: histograms of \dg \, and \dgg \,
 for galaxies selected by $\Delta$SFMS.
 Bottom panels: histograms of $\Delta$SFMS for galaxies selected by
 \dg \, and \dgg.
 Galaxies which are detected in each population are shown as solid
 lines, medians as large solid points, thick lines show
 $\pm$1 standard deviations, and thin lines show the 5th and 95th
 percentiles of each population (detections only). Non-detections in
 each population are shown as dotted  
 histograms. Colored crosses indicate the medians of the full samples
 in each population, including both detections and upper limits of
 non-detections.
    \label{fig:hist4}}
\end{figure*}

\begin{figure*}
\centering
\includegraphics[width=0.735\textwidth]{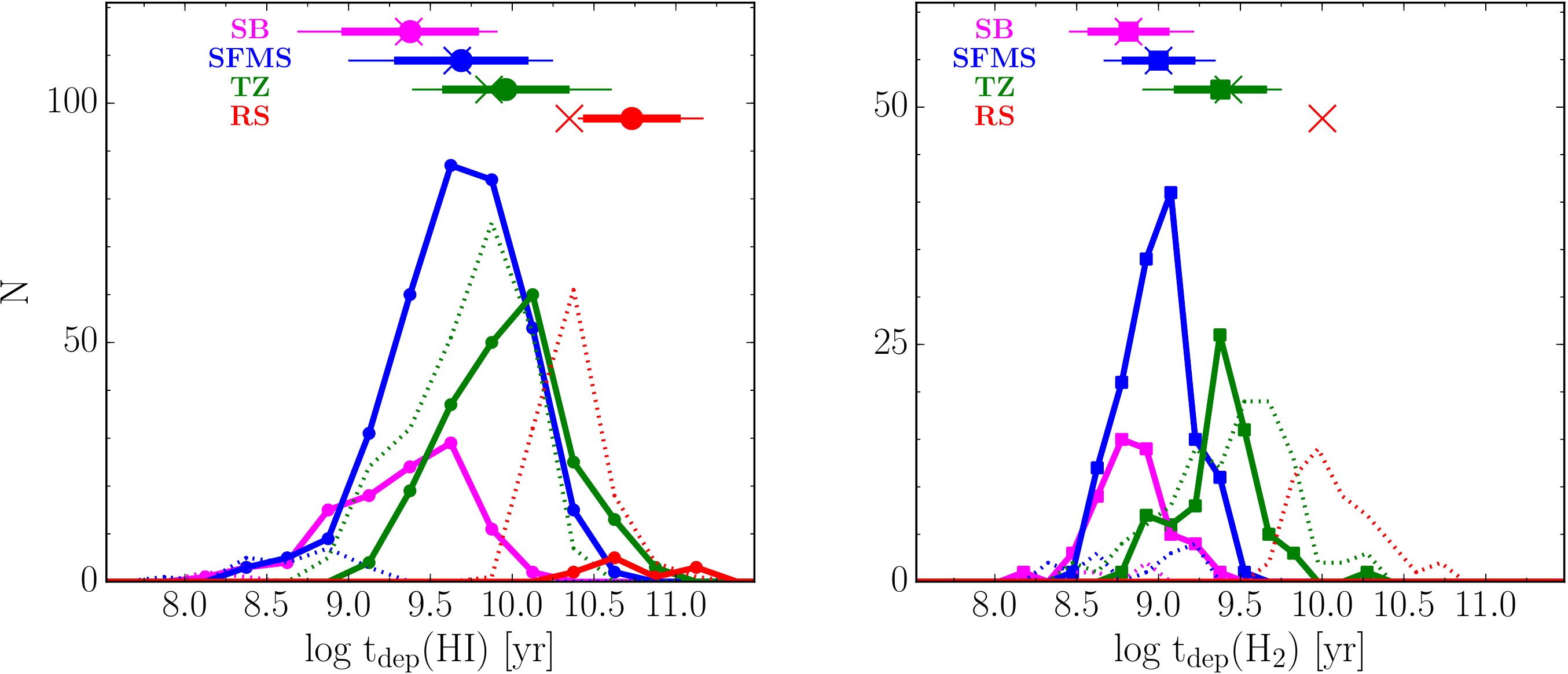}
  \caption{\hi \, (left) and \H2 (right) depletion times for SB
    (magenta), SFMS (blue), and 
    TZ (green) galaxy populations with gas detections (non-detections
    are shown as dotted lines). Medians of the detections in each
    population are shown as large symbols with thick 1~$\sigma$ error bars and
    thin lines extending to the 5th and 95th
    percentiles. Colored crosses show the medians of the full samples
    in each population, including both detections and upper limits of
    non-detections.
    \label{fig:tdep}}
\end{figure*}

\section{Cold gas above/below the SFMS}
\label{sec:off}

Figure~\ref{fig:sfgf} demonstrates the complexity of the
correspondence between star formation and cold gas content. Galaxies
are color coded based on their $\Delta$SFMS from the sSFR-\Mst \, plot
(top two panels). Those within $\pm$0.3~dex (approximately 1$\sigma$)
of the SFMS are shown  
in blue, and those with higher sSFR are referred to as starbursts (SB)
and 
shown in magenta. The Red Sequence (RS) is typically dominant
below sSFR$<$-11.8~yr$^{-1}$ \citep{salim14}, which corresponds to
$\Delta$SFMS$<$-1.55~dex for our SFMS, and is shown in red. Between
the RS and the SFMS is the Transition Zone (TZ), shown in
green. The precise threshold between the TZ and the RS is somewhat
arbitrary; we adopt $\Delta$SFMS$=$-1.55~dex since below that point
only $\sim$10\% (0\%) of the galaxies are detected in \hi \, (CO). As
we are interested in the cold gas properties of TZ galaxies, this RS
definition makes our TZ population more meaningful. Our main
conclusions would remain unchanged if we adopted variations on this
threshold between the TZ and the RS.

On the top left panel of Figure~\ref{fig:sfgf}, filled circles
denote galaxies which have been detected in 
\hi, while colored crosses show sources which do not have
\hi \, detections (although still have valid sSFR measurements). The
top right panel of Figure~\ref{fig:sfgf} shows the same for molecular
gas (CO) observations: filled squares show sources with molecular gas
detections, while colored crosses denote those which have 
been observed but not detected (again, these are valid sSFR measurements).

\begin{table*}
  \centering
	\caption{Population sizes (detections and non-detections),
          averages and standard deviations (of detections only), and
          medians (of detections and non-detections) are given
          for each selection.
          $<$\dg$>$ and $<$\dgg$>$
          are computed for SFMS-selected populations (SB, SFMS, TZ, RS),
          and 
          $<$$\Delta$SFMS$>$ is computed for \dg- and \dgg-selected 
          populations (GR, GN, GP). When fewer than half of the
          galaxies in a selected population are detected, medians are
          not computed.
	  \label{tab:sf2gf}}
        \begin{tabular}{|c|c|r|l|r|c|c|r|l|r|c|}
 & & \multicolumn{4}{c}{\hi} & & \multicolumn{4}{c}{\H2} \\ 
\hline
$\Delta$SFMS & & N$_\textrm{d}$&(N$_\textrm{nd}$) & $<$$\Delta$HIMS$>$ & $<$$\Delta$HIMS$>_\textrm{med}$ & & N$_\textrm{d}$&(N$_\textrm{nd}$) & $<$$\Delta$H$_2$MS$>$ & $<$$\Delta$H$_2$MS$>_\textrm{med}$ \\ 
\hline 
SB & & 107&(3) & $+0.10~(0.35)$ & $+0.13$ &  & 52&(4) & $+0.33~(0.24)$ & $-0.20$ \\ 
SFMS & & 349&(20) & $-0.03~(0.36)$ & $-0.02$ & & 136&(14) & $+0.01~(0.20)$ & $-0.02$ \\ 
TZ & & 211&(246) & $-0.33~(0.36)$ & -- & & 73&(105) & $-0.30~(0.27)$ & -- \\ 
RS & & 11&(117) & $-0.53~(0.28)$ & -- & & 0&(50) & -- & -- \\ 
\hline
$\Delta$HIMS & & N$_\textrm{d}$&(N$_\textrm{nd}$) & $<$$\Delta$SFMS$>$ & $<$$\Delta$SFMS$>_\textrm{med}$ & & N$_\textrm{d}$&(N$_\textrm{nd}$) & $<$$\Delta$SFMS$>$ & $<$$\Delta$SFMS$>_\textrm{med}$ \\ 
\hline 
GR(2) & & 106&(0) & $+0.15~(0.31)$ & $+0.17$ & & 65&(4) & $+0.33~(0.30)$ & $+0.36$ \\ 
GN(2) & & 350&(0) & $-0.08~(0.41)$ & $-0.03$ & & 126&(14) & $-0.00~(0.27)$ & $+0.03$ \\ 
GP(2) & & 222&(386) & $-0.43~(0.55)$ & $-1.02$ & & 70&(155) & $-0.51~(0.35)$ & -- \\ 
\hline 
\end{tabular}
\end{table*}

The bottom two panels of Figure~\ref{fig:sfgf} show the \hi \, and \H2
gas fraction scaling relations, using the $\Delta$SFMS-selected color
scheme. The bottom left panel plots the \hi \, gas fraction scaling
relation, with 
colored circles showing \hi \, detected galaxies and colored crosses at
the upper limits of non-detections. The bottom right panel plots the
\H2 gas fraction scaling relation, with colored squares showing
detections and colored crosses at the upper limits of non-detections.
 Analogously to the SFMS, here we show the GFMS relation (i.e.,
fitting the gas fraction scaling relation in \hi \, and \H2 for SFMS
galaxies only), and identify galaxies within $\pm$0.3~dex (0.2~dex for
\H2) as gas normal (GN). These widths correspond to 1$\sigma$ standard
deviations of the gas fraction distributions of the SFMS
population. Galaxies above this sequence are considered
gas rich (GR) and below are gas poor (GP). 

There is a general correspondence between these two relations in that
the majority of galaxies above the SFMS are also above the 
GFMS \citep[as was also shown in][]{saintonge16}, but there is
substantial scatter, especially in the \hi \, 
relations. In particular, a significant number of TZ galaxies below
the SFMS have \hi \, gas fractions which are consistent with (GN) or
even above (GR) that of SFMS galaxies. 
Before discussing specific trends, we first quantify the
strength of this correspondence between $\Delta$SFMS and \dg.

\subsection{Correspondence between $\Delta$SFMS and \dg}

First we consider the $\Delta$SFMS populations (SB, SFMS, TZ, and RS) and plot
the distribution of \dg \, 
 for each in the top left
panel of Figure~\ref{fig:hist4}. We include galaxies with \hi \,
detections in the solid histogram (for which statistics are computed)
and show the upper limits of non-detections as dotted histograms for
reference. These populations all have broad 
distributions of \hi \, gas fractions with 1$\sigma$ widths of
$\sim$0.4~dex. The average \dg \,
 of the SB, SFMS, and TZ
populations varies by only 0.15-0.3~dex, so there is significant
overlap between galaxies above, on, and below the SFMS.

Likewise the top right panel of Figure~\ref{fig:hist4} shows the
\dgg \, 
distributions for the molecular gas of the same 
populations. Here the 
distributions are much tighter ($\sim$0.25~dex) but their averages are
separated by similar amounts (0.2-0.3~dex). As was visually apparent
in Figure~\ref{fig:sfgf}, there is a tighter correspondence when
comparing SFR with \mH2 than with \mhi. As noted in
\citet{catinella18}, galaxies in the SFMS have a narrower range of \H2
gas fractions, whereas the \hi \, gas fraction can change by two
orders of magnitude. The averages and standard deviations of the \dg
\, and \dgg \, distributions are quantified in Table~\ref{tab:sf2gf}.

Carrying out a similar selection process in the reverse direction, the
bottom left 
panel of Figure~\ref{fig:hist4} shows the $\Delta$SFMS distributions
of the GFMS-selected populations (GR, GN, GP). Here again the
correlation between \hi \, content and SFR is weak, as all three
distributions have significant overlap. The most gas-rich galaxies
have the narrowest distribution of $\Delta$SFMS, suggesting that
elevated star formation is very likely given
a rich supply of atomic gas. However, even this gas-rich population has a
tail that extends to $-$1$\sigma_\textrm{SF}$ below the SFMS.

The bottom right panel of Figure~\ref{fig:hist4} shows the
$\Delta$SFMS distributions for the GF2MS-selected populations (GR2,
GN2, GP2). These distributions are again tighter than those selected
by \dg, 
but not as tight as those in the top right
panel. This suggests that the correlation from SFR to \H2 content
is stronger than the correlation from \H2 content to SFR. The averages
and standard deviations of these $\Delta$SFMS distributions for the
populations selected by \hi \, and \H2 gas fractions are also given in
Table~\ref{tab:sf2gf}.

\subsection{Depletion times and the SFMS}
\label{sec:tdep}

\begin{figure*}
\centering
\includegraphics[width=0.99\textwidth]{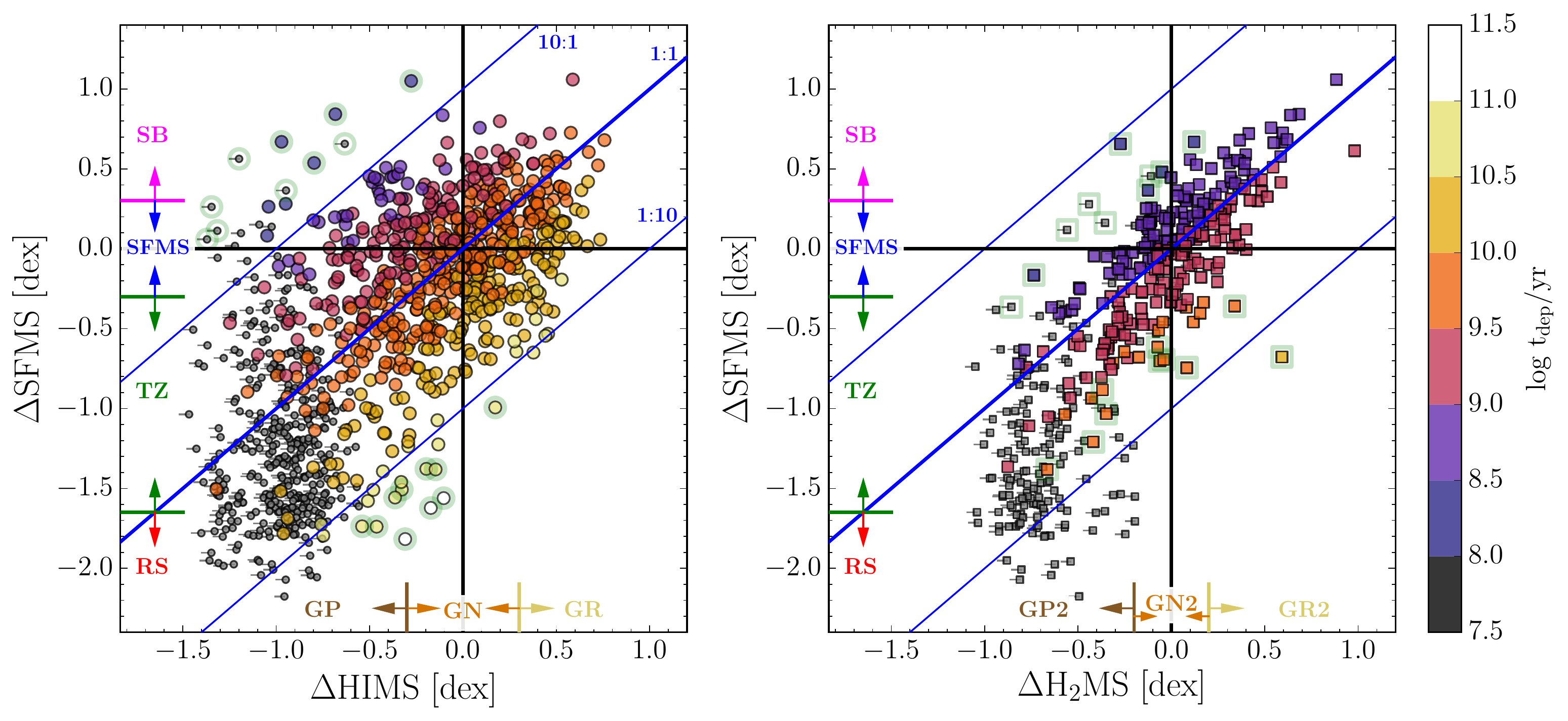}
  \caption{$\Delta$SFMS is plotted vs.~\dg \, and \dgg,
    color coded by depletion time (using circles for \mhi \, and
    squares for \mH2,
     respectively). Regions of $\Delta$SFMS and \dg \,
     populations are indicated along each axis. Galaxies with \hi/CO
     non-detections are shown as 
     small grey symbols at their upper limit \dg \,
   with short lines
     extending toward permitted gas fractions.
    Diagonal lines show 1:1, 10:1, and 1:10 ratios.
    Note that the depletion time at the origin is  
    different between the two panels ($\sim$10$^{9.6}$~yr in \hi, and
    $\sim$10$^{9.0}$~yr in \H2).
    The ten longest and ten shortest depletion time galaxies in each
    panel are surrounded by light green halos and are discussed
    further in Section~\ref{sec:extreme} (see also Figure~\ref{fig:ex}).
    \label{fig:dd}}
\end{figure*}

Galaxies above the SFMS (with higher SFR) are likely to deplete their
cold gas reservoirs more quickly than those below the SFMS (with lower
SFR). However, given the wide scatter in gas properties below the
SFMS, we next compute depletion times
(t$_\textrm{dep}$=M$_\textrm{gas}$/SFR) for each 
galaxy using their \hi \, and \H2 masses. These depletion times
represent the timescale for total gas consumption when assuming a very
simplistic constant SFR and no gas recycling.
Some previous work has
suggested that star-forming galaxies are re-fueled on $\sim$Gyr
timescales on average, which complicates this simple picture
\citep{kannappan13}. 
Furthermore, the physical extents of the \hi \, and molecular
  gas distributions are likely quite different, and our unresolved
  observations cannot distinguish between gas in the far outskirts
  (which must first migrate inwards before it can play a role in star
  formation) and gas in the inner regions (which may be consumed on
  shorter time scales).
 Nonetheless, the \hi \,
and \H2 depletion times are useful to estimate the evolutionary
potential of galaxies on and outside the SFMS.
Previous observations of \H2 depletion time 
have shown a near constant value (0.7~Gyr) in SFMS galaxies across
cosmic time \citep{tacconi13}, while others show a dependence on sSFR
such that more star-forming galaxies have shorter
depletion times \citep{saintonge11,saintonge16}.

\newcommand{\w}{0.095\textwidth}
\newcommand{\ww}{0.090\textwidth}

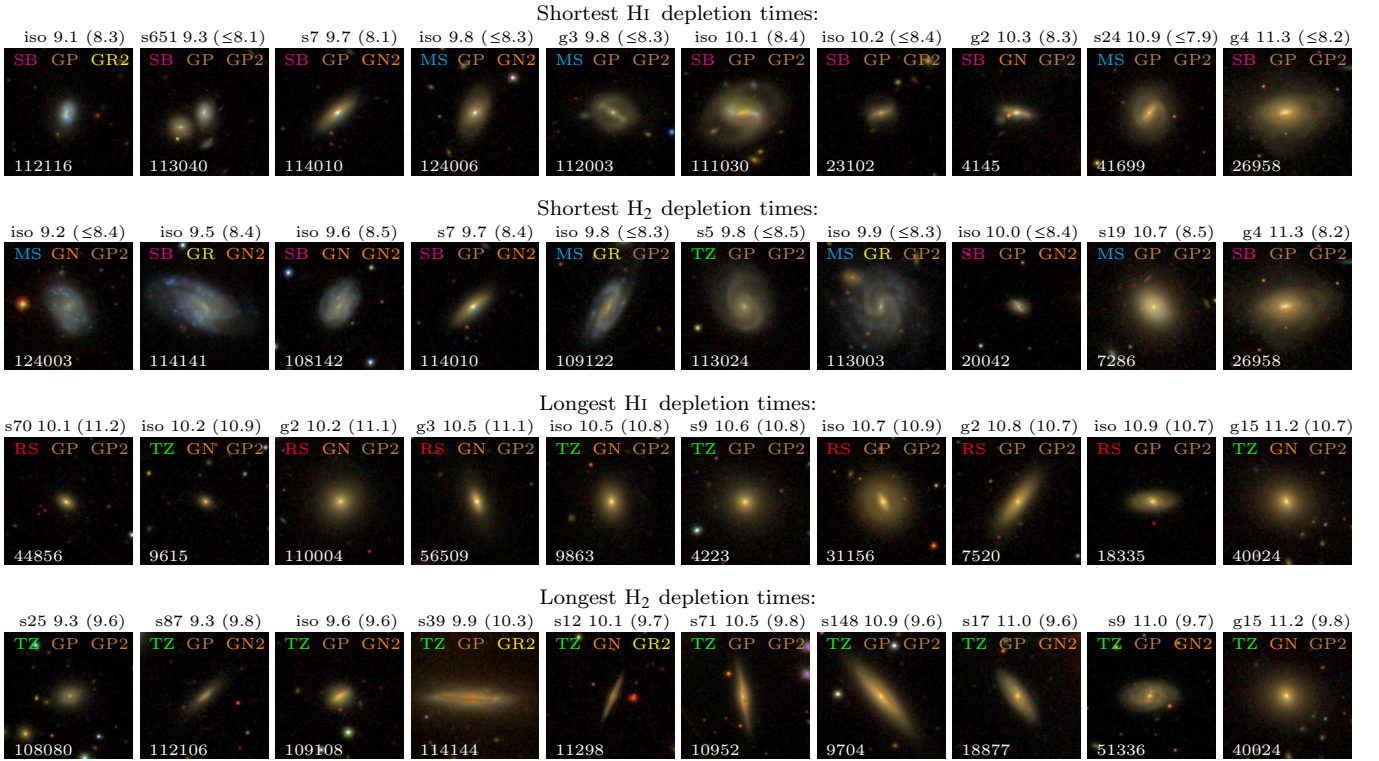
\begin{figure*}
  \centering
  Shortest \hi \, depletion times:\\
\FramedBoxx{\ww}{iso~9.1~(8.3)\\ \textcolor{white}{.}}
\FramedBoxx{\ww}{s651~9.3~($\le$8.1)\\ \textcolor{white}{.}}
\FramedBoxx{\ww}{s7~9.7~(8.1)\\ \textcolor{white}{.}}
\FramedBoxx{\ww}{iso~9.8~($\le$8.3)\\ \textcolor{white}{.}}
\FramedBoxx{\ww}{g3~9.8~($\le$8.3)\\ \textcolor{white}{.}}
\FramedBoxx{\ww}{iso~10.1~(8.4)\\ \textcolor{white}{.}}
\FramedBoxx{\ww}{iso~10.2~($\le$8.4)\\ \textcolor{white}{.}}
\FramedBoxx{\ww}{g2~10.3~(8.3)\\ \textcolor{white}{.}}
\FramedBoxx{\ww}{s24~10.9~($\le$7.9)\\ \textcolor{white}{.}}
\FramedBoxx{\ww}{g4~11.3~($\le$8.2)\\ \textcolor{white}{.}}
\begin{overpic}[width=\w]{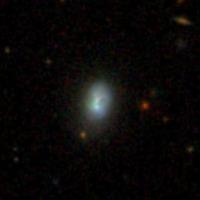}
\put(7,87){\scriptsize\textcolor{magenta}{SB}}
\put(36,87){\scriptsize\textcolor{brown}{GP}}
\put(67,87){\scriptsize\textcolor{yellow}{GR2}}
\put(7,3){\scriptsize\textcolor{white}{112116}}
\end{overpic}
\begin{overpic}[width=\w]{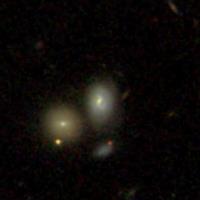}
\put(7,87){\scriptsize\textcolor{magenta}{SB}}
\put(36,87){\scriptsize\textcolor{brown}{GP}}
\put(67,87){\scriptsize\textcolor{brown}{GP2}}
\put(7,3){\scriptsize\textcolor{white}{113040}}
\end{overpic}
\begin{overpic}[width=\w]{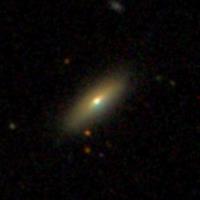}
\put(7,87){\scriptsize\textcolor{magenta}{SB}}
\put(36,87){\scriptsize\textcolor{brown}{GP}}
\put(67,87){\scriptsize\textcolor{orange}{GN2}}
\put(7,3){\scriptsize\textcolor{white}{114010}}
\end{overpic}
\begin{overpic}[width=\w]{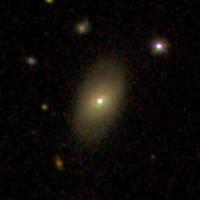}
\put(7,87){\scriptsize\textcolor{cyan}{MS}}
\put(36,87){\scriptsize\textcolor{brown}{GP}}
\put(67,87){\scriptsize\textcolor{orange}{GN2}}
\put(7,3){\scriptsize\textcolor{white}{124006}}
\end{overpic}
\begin{overpic}[width=\w]{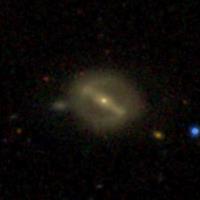}
\put(7,87){\scriptsize\textcolor{cyan}{MS}}
\put(36,87){\scriptsize\textcolor{brown}{GP}}
\put(67,87){\scriptsize\textcolor{brown}{GP2}}
\put(7,3){\scriptsize\textcolor{white}{112003}}
\end{overpic}
\begin{overpic}[width=\w]{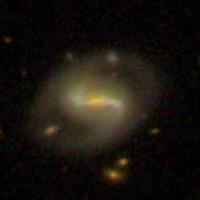}
\put(7,87){\scriptsize\textcolor{magenta}{SB}}
\put(36,87){\scriptsize\textcolor{brown}{GP}}
\put(67,87){\scriptsize\textcolor{brown}{GP2}}
\put(7,3){\scriptsize\textcolor{white}{111030}}
\end{overpic}
\begin{overpic}[width=\w]{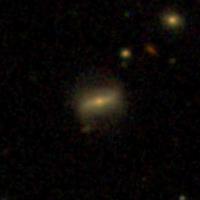}
\put(7,87){\scriptsize\textcolor{magenta}{SB}}
\put(36,87){\scriptsize\textcolor{brown}{GP}}
\put(67,87){\scriptsize\textcolor{brown}{GP2}}
\put(7,3){\scriptsize\textcolor{white}{23102}}
\end{overpic}
\begin{overpic}[width=\w]{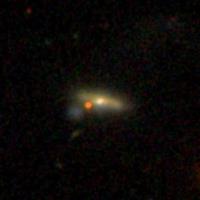}
\put(7,87){\scriptsize\textcolor{magenta}{SB}}
\put(36,87){\scriptsize\textcolor{orange}{GN}}
\put(67,87){\scriptsize\textcolor{brown}{GP2}}
\put(7,3){\scriptsize\textcolor{white}{4145}}
\end{overpic}
\begin{overpic}[width=\w]{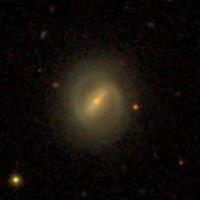}
\put(7,87){\scriptsize\textcolor{cyan}{MS}}
\put(36,87){\scriptsize\textcolor{brown}{GP}}
\put(67,87){\scriptsize\textcolor{brown}{GP2}}
\put(7,3){\scriptsize\textcolor{white}{41699}}
\end{overpic}
\begin{overpic}[width=\w]{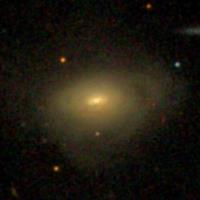}
\put(7,87){\scriptsize\textcolor{magenta}{SB}}
\put(36,87){\scriptsize\textcolor{brown}{GP}}
\put(67,87){\scriptsize\textcolor{brown}{GP2}}
\put(7,3){\scriptsize\textcolor{white}{26958}}
\end{overpic}\\
  \vspace{5pt}
  Shortest \H2 depletion times:\\
  \FramedBoxx{\ww}{iso~9.2~($\le$8.4)\\ \textcolor{white}{.}}
\FramedBoxx{\ww}{iso~9.5~(8.4)\\ \textcolor{white}{.}}
\FramedBoxx{\ww}{iso~9.6~(8.5)\\ \textcolor{white}{.}}
\FramedBoxx{\ww}{s7~9.7~(8.4)\\ \textcolor{white}{.}}
\FramedBoxx{\ww}{iso~9.8~($\le$8.3)\\ \textcolor{white}{.}}
\FramedBoxx{\ww}{s5~9.8~($\le$8.5)\\ \textcolor{white}{.}}
\FramedBoxx{\ww}{iso~9.9~($\le$8.3)\\ \textcolor{white}{.}}
\FramedBoxx{\ww}{iso~10.0~($\le$8.4)\\ \textcolor{white}{.}}
\FramedBoxx{\ww}{s19~10.7~(8.5)\\ \textcolor{white}{.}}
\FramedBoxx{\ww}{g4~11.3~(8.2)\\ \textcolor{white}{.}}
\begin{overpic}[width=\w]{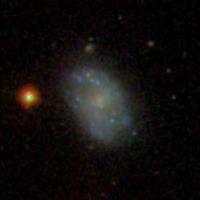}
\put(7,87){\scriptsize\textcolor{cyan}{MS}}
\put(36,87){\scriptsize\textcolor{orange}{GN}}
\put(67,87){\scriptsize\textcolor{brown}{GP2}}
\put(7,3){\scriptsize\textcolor{white}{124003}}
\end{overpic}
\begin{overpic}[width=\w]{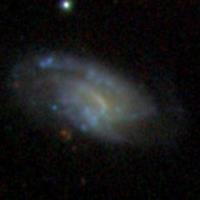}
\put(7,87){\scriptsize\textcolor{magenta}{SB}}
\put(36,87){\scriptsize\textcolor{yellow}{GR}}
\put(67,87){\scriptsize\textcolor{orange}{GN2}}
\put(7,3){\scriptsize\textcolor{white}{114141}}
\end{overpic}
\begin{overpic}[width=\w]{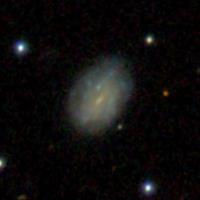}
\put(7,87){\scriptsize\textcolor{magenta}{SB}}
\put(36,87){\scriptsize\textcolor{orange}{GN}}
\put(67,87){\scriptsize\textcolor{orange}{GN2}}
\put(7,3){\scriptsize\textcolor{white}{108142}}
\end{overpic}
\begin{overpic}[width=\w]{114010_sdss_small2.jpg}
\put(7,87){\scriptsize\textcolor{magenta}{SB}}
\put(36,87){\scriptsize\textcolor{brown}{GP}}
\put(67,87){\scriptsize\textcolor{orange}{GN2}}
\put(7,3){\scriptsize\textcolor{white}{114010}}
\end{overpic}
\begin{overpic}[width=\w]{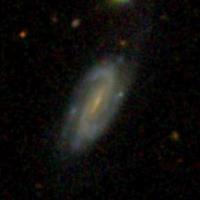}
\put(7,87){\scriptsize\textcolor{cyan}{MS}}
\put(36,87){\scriptsize\textcolor{yellow}{GR}}
\put(67,87){\scriptsize\textcolor{brown}{GP2}}
\put(7,3){\scriptsize\textcolor{white}{109122}}
\end{overpic}
\begin{overpic}[width=\w]{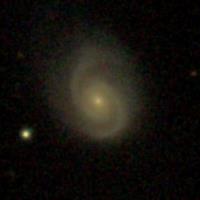}
\put(7,87){\scriptsize\textcolor{green}{TZ}}
\put(36,87){\scriptsize\textcolor{brown}{GP}}
\put(67,87){\scriptsize\textcolor{brown}{GP2}}
\put(7,3){\scriptsize\textcolor{white}{113024}}
\end{overpic}
\begin{overpic}[width=\w]{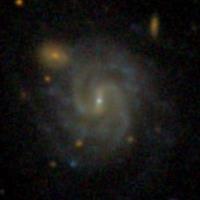}
\put(7,87){\scriptsize\textcolor{cyan}{MS}}
\put(36,87){\scriptsize\textcolor{yellow}{GR}}
\put(67,87){\scriptsize\textcolor{brown}{GP2}}
\put(7,3){\scriptsize\textcolor{white}{113003}}
\end{overpic}
\begin{overpic}[width=\w]{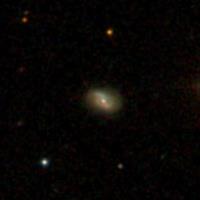}
\put(7,87){\scriptsize\textcolor{magenta}{SB}}
\put(36,87){\scriptsize\textcolor{brown}{GP}}
\put(67,87){\scriptsize\textcolor{orange}{GN2}}
\put(7,3){\scriptsize\textcolor{white}{20042}}
\end{overpic}
\begin{overpic}[width=\w]{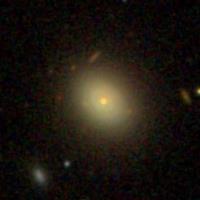}
\put(7,87){\scriptsize\textcolor{cyan}{MS}}
\put(36,87){\scriptsize\textcolor{brown}{GP}}
\put(67,87){\scriptsize\textcolor{brown}{GP2}}
\put(7,3){\scriptsize\textcolor{white}{7286}}
\end{overpic}
\begin{overpic}[width=\w]{26958_sdss_small2.jpg}
\put(7,87){\scriptsize\textcolor{magenta}{SB}}
\put(36,87){\scriptsize\textcolor{brown}{GP}}
\put(67,87){\scriptsize\textcolor{brown}{GP2}}
\put(7,3){\scriptsize\textcolor{white}{26958}}
\end{overpic}%
  \vspace{5pt}
  Longest \hi \, depletion times:\\
\FramedBoxx{\ww}{s70~10.1~(11.2)\\ \textcolor{white}{.}}
\FramedBoxx{\ww}{iso~10.2~(10.9)\\ \textcolor{white}{.}}
\FramedBoxx{\ww}{g2~10.2~(11.1)\\ \textcolor{white}{.}}
\FramedBoxx{\ww}{g3~10.5~(11.1)\\ \textcolor{white}{.}}
\FramedBoxx{\ww}{iso~10.5~(10.8)\\ \textcolor{white}{.}}
\FramedBoxx{\ww}{s9~10.6~(10.8)\\ \textcolor{white}{.}}
\FramedBoxx{\ww}{iso~10.7~(10.9)\\ \textcolor{white}{.}}
\FramedBoxx{\ww}{g2~10.8~(10.7)\\ \textcolor{white}{.}}
\FramedBoxx{\ww}{iso~10.9~(10.7)\\ \textcolor{white}{.}}
\FramedBoxx{\ww}{g15~11.2~(10.7)\\ \textcolor{white}{.}}
\begin{overpic}[width=\w]{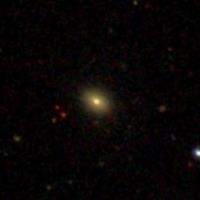}
\put(7,87){\scriptsize\textcolor{red}{RS}}
\put(36,87){\scriptsize\textcolor{brown}{GP}}
\put(67,87){\scriptsize\textcolor{brown}{GP2}}
\put(7,3){\scriptsize\textcolor{white}{44856}}
\end{overpic}
\begin{overpic}[width=\w]{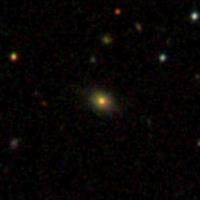}
\put(7,87){\scriptsize\textcolor{green}{TZ}}
\put(36,87){\scriptsize\textcolor{orange}{GN}}
\put(67,87){\scriptsize\textcolor{brown}{GP2}}
\put(7,3){\scriptsize\textcolor{white}{9615}}
\end{overpic}
\begin{overpic}[width=\w]{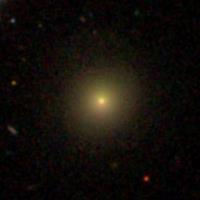}
\put(7,87){\scriptsize\textcolor{red}{RS}}
\put(36,87){\scriptsize\textcolor{orange}{GN}}
\put(67,87){\scriptsize\textcolor{brown}{GP2}}
\put(7,3){\scriptsize\textcolor{white}{110004}}
\end{overpic}
\begin{overpic}[width=\w]{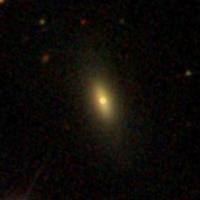}
\put(7,87){\scriptsize\textcolor{red}{RS}}
\put(36,87){\scriptsize\textcolor{orange}{GN}}
\put(67,87){\scriptsize\textcolor{brown}{GP2}}
\put(7,3){\scriptsize\textcolor{white}{56509}}
\end{overpic}
\begin{overpic}[width=\w]{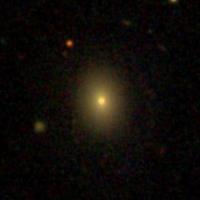}
\put(7,87){\scriptsize\textcolor{green}{TZ}}
\put(36,87){\scriptsize\textcolor{orange}{GN}}
\put(67,87){\scriptsize\textcolor{brown}{GP2}}
\put(7,3){\scriptsize\textcolor{white}{9863}}
\end{overpic}
\begin{overpic}[width=\w]{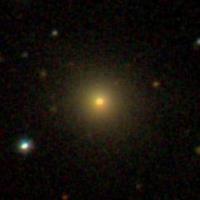}
\put(7,87){\scriptsize\textcolor{green}{TZ}}
\put(36,87){\scriptsize\textcolor{brown}{GP}}
\put(67,87){\scriptsize\textcolor{brown}{GP2}}
\put(7,3){\scriptsize\textcolor{white}{4223}}
\end{overpic}
\begin{overpic}[width=\w]{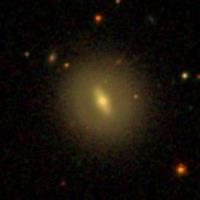}
\put(7,87){\scriptsize\textcolor{red}{RS}}
\put(36,87){\scriptsize\textcolor{brown}{GP}}
\put(67,87){\scriptsize\textcolor{brown}{GP2}}
\put(7,3){\scriptsize\textcolor{white}{31156}}
\end{overpic}
\begin{overpic}[width=\w]{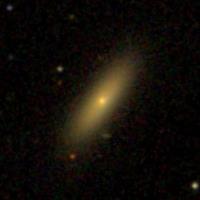}
\put(7,87){\scriptsize\textcolor{red}{RS}}
\put(36,87){\scriptsize\textcolor{brown}{GP}}
\put(67,87){\scriptsize\textcolor{brown}{GP2}}
\put(7,3){\scriptsize\textcolor{white}{7520}}
\end{overpic}
\begin{overpic}[width=\w]{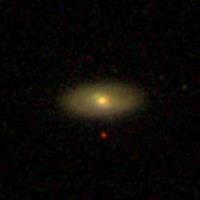}
\put(7,87){\scriptsize\textcolor{red}{RS}}
\put(36,87){\scriptsize\textcolor{brown}{GP}}
\put(67,87){\scriptsize\textcolor{brown}{GP2}}
\put(7,3){\scriptsize\textcolor{white}{18335}}
\end{overpic}
\begin{overpic}[width=\w]{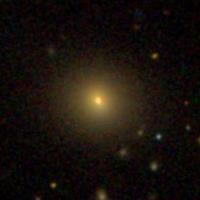}
\put(7,87){\scriptsize\textcolor{green}{TZ}}
\put(36,87){\scriptsize\textcolor{orange}{GN}}
\put(67,87){\scriptsize\textcolor{brown}{GP2}}
\put(7,3){\scriptsize\textcolor{white}{40024}}
\end{overpic}\\
  \vspace{5pt}
  Longest \H2 depletion times:\\
\FramedBoxx{\ww}{s25~9.3~(9.6)\\ \textcolor{white}{.}}
\FramedBoxx{\ww}{s87~9.3~(9.8)\\ \textcolor{white}{.}}
\FramedBoxx{\ww}{iso~9.6~(9.6)\\ \textcolor{white}{.}}
\FramedBoxx{\ww}{s39~9.9~(10.3)\\ \textcolor{white}{.}}
\FramedBoxx{\ww}{s12~10.1~(9.7)\\ \textcolor{white}{.}}
\FramedBoxx{\ww}{s71~10.5~(9.8)\\ \textcolor{white}{.}}
\FramedBoxx{\ww}{s148~10.9~(9.6)\\ \textcolor{white}{.}}
\FramedBoxx{\ww}{s17~11.0~(9.6)\\ \textcolor{white}{.}}
\FramedBoxx{\ww}{s9~11.0~(9.7)\\ \textcolor{white}{.}}
\FramedBoxx{\ww}{g15~11.2~(9.8)\\ \textcolor{white}{.}}
\begin{overpic}[width=\w]{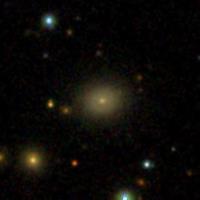}
\put(7,87){\scriptsize\textcolor{green}{TZ}}
\put(36,87){\scriptsize\textcolor{brown}{GP}}
\put(67,87){\scriptsize\textcolor{brown}{GP2}}
\put(7,3){\scriptsize\textcolor{white}{108080}}
\end{overpic}
\begin{overpic}[width=\w]{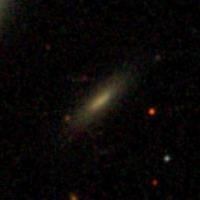}
\put(7,87){\scriptsize\textcolor{green}{TZ}}
\put(36,87){\scriptsize\textcolor{brown}{GP}}
\put(67,87){\scriptsize\textcolor{orange}{GN2}}
\put(7,3){\scriptsize\textcolor{white}{112106}}
\end{overpic}
\begin{overpic}[width=\w]{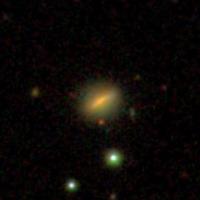}
\put(7,87){\scriptsize\textcolor{green}{TZ}}
\put(36,87){\scriptsize\textcolor{brown}{GP}}
\put(67,87){\scriptsize\textcolor{orange}{GN2}}
\put(7,3){\scriptsize\textcolor{white}{109108}}
\end{overpic}
\begin{overpic}[width=\w]{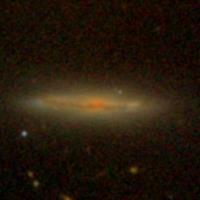}
\put(7,87){\scriptsize\textcolor{green}{TZ}}
\put(36,87){\scriptsize\textcolor{brown}{GP}}
\put(67,87){\scriptsize\textcolor{yellow}{GR2}}
\put(7,3){\scriptsize\textcolor{white}{114144}}
\end{overpic}
\begin{overpic}[width=\w]{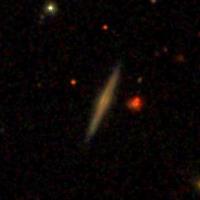}
\put(7,87){\scriptsize\textcolor{green}{TZ}}
\put(36,87){\scriptsize\textcolor{orange}{GN}}
\put(67,87){\scriptsize\textcolor{yellow}{GR2}}
\put(7,3){\scriptsize\textcolor{white}{11298}}
\end{overpic}
\begin{overpic}[width=\w]{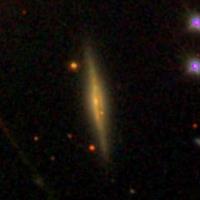}
\put(7,87){\scriptsize\textcolor{green}{TZ}}
\put(36,87){\scriptsize\textcolor{brown}{GP}}
\put(67,87){\scriptsize\textcolor{brown}{GP2}}
\put(7,3){\scriptsize\textcolor{white}{10952}}
\end{overpic}
\begin{overpic}[width=\w]{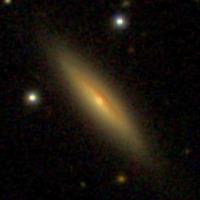}
\put(7,87){\scriptsize\textcolor{green}{TZ}}
\put(36,87){\scriptsize\textcolor{brown}{GP}}
\put(67,87){\scriptsize\textcolor{brown}{GP2}}
\put(7,3){\scriptsize\textcolor{white}{9704}}
\end{overpic}
\begin{overpic}[width=\w]{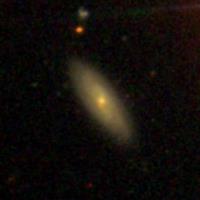}
\put(7,87){\scriptsize\textcolor{green}{TZ}}
\put(36,87){\scriptsize\textcolor{brown}{GP}}
\put(67,87){\scriptsize\textcolor{orange}{GN2}}
\put(7,3){\scriptsize\textcolor{white}{18877}}
\end{overpic}
\begin{overpic}[width=\w]{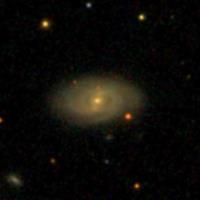}
\put(7,87){\scriptsize\textcolor{green}{TZ}}
\put(36,87){\scriptsize\textcolor{brown}{GP}}
\put(67,87){\scriptsize\textcolor{orange}{GN2}}
\put(7,3){\scriptsize\textcolor{white}{51336}}
\end{overpic}
\begin{overpic}[width=\w]{40024_sdss_small2.jpg}
\put(7,87){\scriptsize\textcolor{green}{TZ}}
\put(36,87){\scriptsize\textcolor{orange}{GN}}
\put(67,87){\scriptsize\textcolor{brown}{GP2}}
\put(7,3){\scriptsize\textcolor{white}{40024}}
\end{overpic}\\
  \label{fig:ex1}
  \caption{SDSS thumbnail images ($80$$''$ on a side) are shown of
    galaxies with the extreme \hi \, and \H2 depletion times, ordered
    by increasing stellar mass from left to right. Above
    each thumbnail is its environmental identity (\textbf{iso}lated
    central or \textbf{s}atellite/\textbf{g}roup central with
    the number of group members), the logarithm of its stellar mass
    (in solar units) and the logarithm of its depletion time (in years
    and in parentheses). Overlaid on each image are its $\Delta$SFMS,
    \dg, and \dgg \, 
   categories, and its xGASS ID.
   \label{fig:ex} }
\end{figure*}

In this work we use the depletion
time as a crude estimate of the evolutionary pace of a galaxy: those
with shorter depletion times are likely to be more rapidly evolving
and transforming
(i.e., consuming a significant fraction of their gas supply on short
timescales), while those with longer depletion times are evolving more
slowly. 
The left panel of Figure~\ref{fig:tdep} shows histograms of \hi \,
depletion times for each $\Delta$SFMS population (SB, SFMS, TZ, RS)
using the same color-coding as previous figures. Within each
population, average \hi \,  depletion times based on \hi \, detections
are shown with large symbols with 1-$\sigma$ error bars as well as 5th
and 95th percentiles. Crosses show medians within each population,
using both detections and non-detections.

Galaxies in the xGASS SFMS (in blue on Figure~\ref{fig:tdep})
span a wide range of \hi \, depletion
times varying from 
0.8$-$40~Gyr. The average \hi \, depletion time is
$\sim$6.2~Gyr with a scatter of 0.6~dex.   
In the TZ below the SFMS (plotted in green), 
galaxies
span a similar dynamic range of \hi \, depletion times (from
7$-$70~Gyr, one order of magnitude), with an
average of 
$\sim$9.7~Gyr. Above the SFMS, SB galaxies (plotted in magenta) 
again have a larger dynamic range (1.3$-$130~Gyr) but a shorter average \hi \,
depletion time of
$\sim$2.9~Gyr.

The right panel of Figure~\ref{fig:tdep} shows histograms of \H2
depletion times for the same $\Delta$SFMS populations with the same
color-coding. Within the SFMS,
the dynamic range of
\H2 depletion times is similar that in \hi, but offset to shorter 
times, as shown in the 
right panel of Figure~\ref{fig:tdep}, spanning only 
0.3$-$3~Gyr, with an average of 
$\sim$1.1~Gyr and a scatter of 0.49~dex. As with \hi, SB galaxies have 
shorter \H2 depletion times (average of 0.8~Gyr), 
and galaxies in the TZ have longer \H2 depletion times (average of 2.8~Gyr). 

Combining all of the cold gas and SFR information, we plot 
$\Delta$SFMS vs.~\dg \, and \dgg \, 
in Figure~\ref{fig:dd}, in a similar 
way to Figure~11 of \citet{schiminovich10}. These plots 
show the general trend for galaxies above (below) the average \hi \, 
and \H2 scaling relations to have positive (negative) $\Delta$SFMS, 
and they also demonstrate the significant amount of scatter in this 
trend. Points in Figure~\ref{fig:dd} are color coded by their gas 
depletion time, and SFMS- and GFMS-selected populations are labeled
along the axes.

\subsection{Galaxies with extreme depletion times}
\label{sec:extreme}

Given the broad correspondence between \dg \,
and $\Delta$SFMS
in Figure~\ref{fig:dd}, we take a closer look at those which deviate
from this general trend. That is, we examine galaxies with the most
extreme depletion times to explore what factors may contribute to
their departure from the main relation. It is important to note that
a galaxy's SFR is expected to stochastically fluctuate, so a galaxy
which currently exhibits an extreme depletion time may be experiencing
a brief fluctuation in its SFR (increase or decrease) and not a
permanent transition. These fluctuations can have particularly strong
effects on the lowest mass galaxies in our sample.

First, we select the 10 galaxies with the shortest \hi \, depletion
times and indicate their locations with light green halos in
the top left corner of the left panel of Figure~\ref{fig:dd}. We
include \hi \, non-detections in this selection, as these translate
directly to upper limits on depletion times, and note that these short
\hi \, depletion time galaxies are all in the SFMS or SB
populations. SDSS thumbnail images are shown for these galaxies in the
top row of Figure~\ref{fig:ex}. We also include the environmental
identity for each galaxy from the \citet{yang07} DR7 group catalog
(discussed further in Section~\ref{sec:gv}), and do not include the 16
galaxies from our sample which lack environmental identities largely
due to their proximity to survey edges.

These short \hi \, depletion time galaxies have (potentially
completely) diminished gas reservoirs but 
unexpectedly strong star formation. As seen in their
SDSS thumbnails, at least half show strong bars or ring features,
suggesting they may have more efficient star formation than an
otherwise similar galaxy without a bar. These short depletion time
galaxies are likely to be the most rapidly evolving systems. Without
an additional supply of gas, their SFR will soon decrease and they may
drop back down to (or even below) the 1:1 line on
Figure~\ref{fig:dd}.

Similarly, we select the 10 galaxies with the shortest \H2 depletion
times and indicate their positions in the right panel of
Figure~\ref{fig:dd}. These are found largely in the SB and
SFMS, but one is in the TZ. Thumbnails from SDSS of this population
are shown in the second row of Figure~\ref{fig:ex}, and they have
similar morphologies to the galaxies with the longest \hi \, depletion
times. Only two galaxies 
(xGASS~114010 and 26958) are common to both categories: xGASS~114010
has a strong bulge and is a satellite in a group of N=7; xGASS~26985
is a central in a group of N=4 and shows a dramatic system of
shells/rings suggesting a recent interaction.

\begin{figure*}
  \centering
  \includegraphics[width=0.99\textwidth]{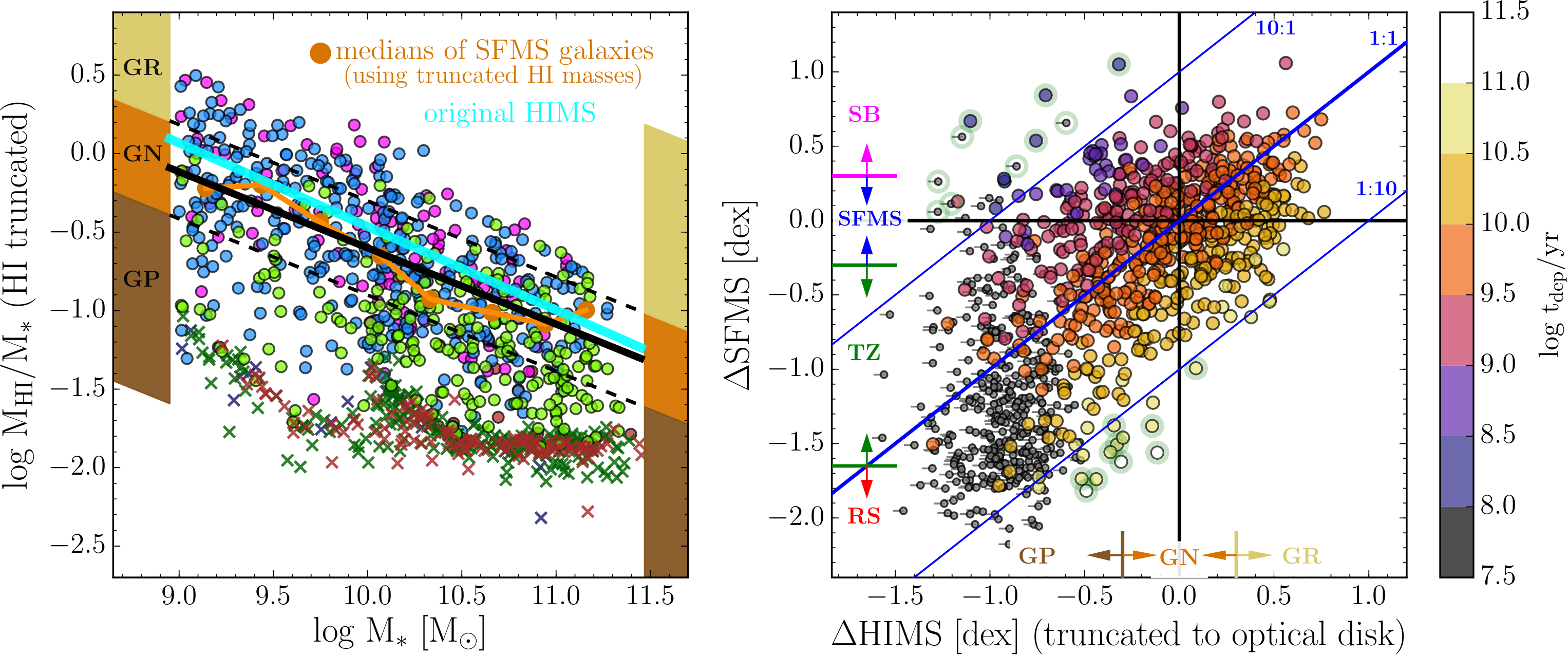}
  \caption{Left panel: \hi \, gas-fraction scaling relation is plotted
    using \hi \, masses truncated within the the optical disk, as
    described in Section~\ref{sec:trunc}. Orange
    points and black line indicate the median relation for SFMS
    galaxies, compared to the original relation from
    Figure~\ref{fig:sfgf} shown in cyan.
    Right panel: truncated \hi \, masses are used to reproduce the
    left panel of Figure~\ref{fig:dd}, using the updated HIMS from the
    left panel. While
    subtle, there is a small systematic offset towards lower values of
    \dg, but the same appearance overall.
    \label{fig:trunc}}
\end{figure*}

At the other extreme, we consider the 10 galaxies with the
longest \hi \, depletion times, which are also indicated with light
green halos in the bottom right corner of the left panel of
Figure~\ref{fig:dd}, and their thumbnails are shown in the third row
of Figure~\ref{fig:ex}. Here we require \hi \, detections to compute
meaningful depletion times, as upper limits on \hi \, masses also
yield upper limits on \hi \, depletion times. Without a detection, the
true depletion time could be significantly smaller and much less
extreme. This population of long \hi \, depletion
time galaxies (with good \hi \, detections) is found in the TZ and
RS.

These galaxies have a significantly larger \hi
\,  reservoir than expected from their SFR, so the atomic gas may be
inaccessible, lingering below a  density threshold, or experiencing
too much shear or dispersion to trigger
star formation. This inaccessibility could mean that the gas has arrived
with high angular momentum from a minor merger, so has not yet
collapsed enough to form a proportionate amount of star
formation. Our \hi \, observations are unresolved, so the neutral gas
has not been localized within the galaxy, and may be significantly
more extended than the stellar components
\citep[perhaps analogously to the sample identified and observed 
  by][]{gereb16,gereb18}, but interferometric
 observations of their \hi \, would be required to confirm this
 scenario. 
These galaxies are all fairly massive (\Mst$\ge$$10^{10}$\msun)
and have strong bulges, but at least four show (hints of) disks as
well. Galaxies 
with substantial \hi \, reservoirs like these are candidates for
possible rejuvenation (i.e., a return to the SFMS) if they are able to
access their full gas supply for star formation.
At their current star formation rates, these long-depletion-time
galaxies are some of the most slowly evolving galaxies. They are
unlikely to move significantly within Figure~\ref{fig:dd}
and represent a moderately static gas-rich population.

Analogously, the longest \H2 depletion time population is selected in
the same way and displayed in the 4th row of
Figure~\ref{fig:ex}. Note in particular that GASS~40024 has both an
extreme \hi \, depletion time and an extreme \H2 depletion time, and
is a satellite in a group of N=15 galaxies.
This long \H2 depletion time population also includes galaxies at lower
masses and are all found in the TZ. Here the number of edge-on 
galaxies with visible absorption from dust lanes suggests that some
members of this population may have under-estimated SFRs. Our SFR is
based on both the direct UV emission from young stars and the
re-processed IR emission from dust absorption, but the
edge-on systems with strongly visible dust lanes may be more extincted
than our SFR calibration assumes. If the SFR estimates of edge-on
galaxies are under-estimated, then their depletion times will be
artificially increased.

\subsection{\added{Predicting gas location from scaling relations and
    predicting \hi \, masses within optical disk}}
\label{sec:trunc}

\added{One might consider the role that the spatial location of cold
  gas within the galaxies in our sample plays in determining their
  star formation histories and depletion times. For example, gas-rich TZ
  galaxies may remain gas-rich (with long depletion times) if their
  remaining gas is largely 
  located in their outskirts and only slowly migrates to the inner
  regions to participate in active star formation. Accounting for
  potential differences in \hi \, and stellar distributions would
  require resolved \hi \, observations, but even without those we can
  employ simple scaling relations to predict the amount of \hi \, mass
  contained within the optical disk. In this way we can estimate the
  depletion time considering only the gas which is predicted to be
  contained within the optical disk.
}

\added{We use the \hi \, size-mass relation from \citet{wang16} to
  predict an \hi \, size using only the \hi \, mass of each galaxy in
  our sample. We also adopt a simple exponential profile for the
  radial \hi \, mass distribution. Using the 90\% Petrosian radius
  from the SDSS $r$ image, we integrate the exponential profile within
  this truncation radius to determine the \hi \, mass contained within
  the optical disk. This is a simple approach and relies on the
  assumption that these galaxies all follow the \hi \, size-mass
  relation and all have the same profile shape. Still, these simple
  truncated \hi \, masses will give an indication of the potential
  impact that the spatial distribution of \hi \, could have on our
  results.
}

\added{The left panel of Figure~\ref{fig:trunc} shows an updated
  version of the \hi \, gas fraction scaling relation, originally
  presented in Figure~\ref{fig:sfgf}, but now using the \hi \, masses
  truncated within the optical disk. As expected, this overall
  reduction in \hi \, masses results in a slightly lower median
  relation than with the full \hi \, masses, with a very weak change
  in slope. The right panel of
  Figure~\ref{fig:trunc} plots $\Delta$SFMS  vs \dg \, for these
  truncated \hi \, masses. When compared with the left panel of
  Figure~\ref{fig:dd}, no significant differences are observed. Since
  the truncated \hi \, masses are always smaller than the total \hi \,
  masses, there is again a systematic shift toward lower values of
  \dg. This   does not indicate that the spatial distribution of the
  gas is unimportant, but rather suggests that there are no
  significant effects revealed by this truncation of an assumed \hi \,
  profile within the optical disk. Resolved observations of the \hi \,
  gas would be required to advance this question further.
}

\section{Un-quenched galaxy populations below the SFMS}
\label{sec:gv}

Using the populations described in Figure~\ref{fig:dd}, we next
focus on the galaxies found in the TZ between the SFMS
and the RS. Here we refer to ``quenched'' galaxies as those with 
SFR measurements low enough to be
consistent with no active star formation and such small cold gas
reservoirs as to be undetectable. 
We will consider the types of evolutionary scenarios and
pathways that TZ galaxies
may follow. First, however, it is important to note that galaxies in
the TZ with cold gas detections have on average $\sim$0.35~dex longer \hi \,
and \H2 depletion times than those in the SFMS. This is visible in
Figure~\ref{fig:tdep}, which also makes clear the significant
overlap between both populations.
Of the $\sim$400 galaxies in the TZ, $\sim$50\% are detected in \hi \,
and 40\% of these \hi \, detections are comparably or more gas-rich than
typical SFMS galaxies (i.e., are GN or GR). This 
suggests that a significant fraction of the TZ galaxies are not
actively quenching or rapidly evolving towards the RS. The remaining
TZ galaxies without \hi \, detections 
may be a more rapidly evolving/quenching population, but the
TZ galaxies with \hi \, 
detections can sustain their current SFR longer than galaxies
currently in the SFMS!

\begin{figure*}
  \centering
  \includegraphics[width=0.99\textwidth]{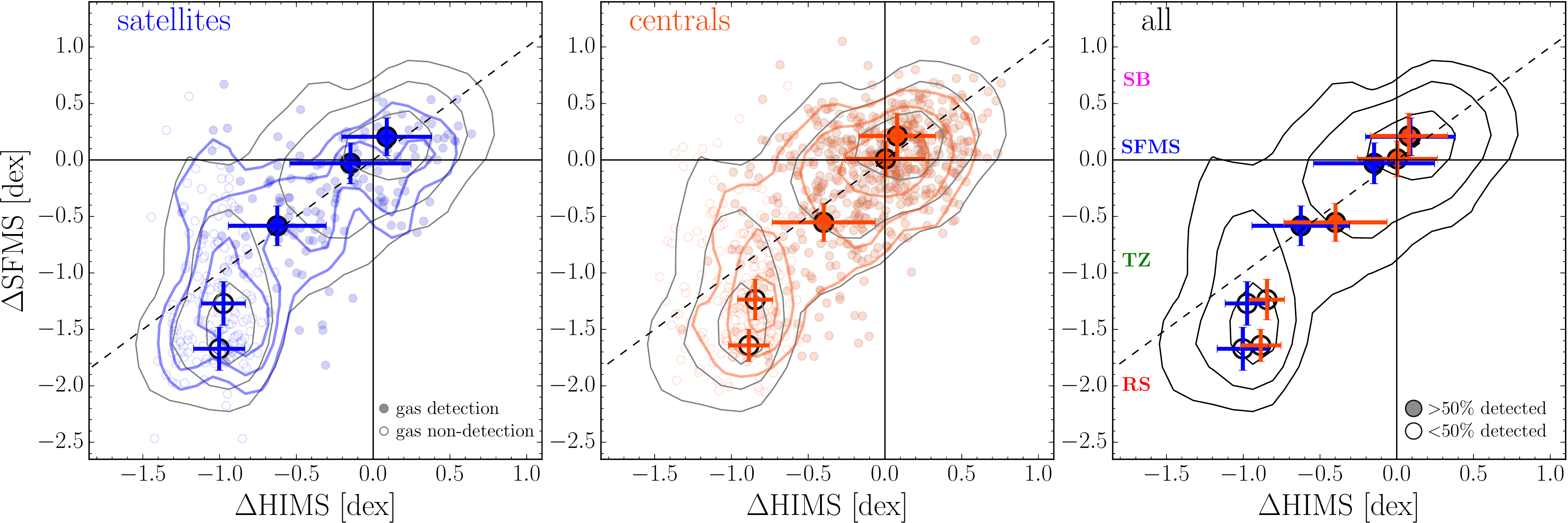}
  \caption{Each panel shows \dg \,
    vs.~$\Delta$SFMS, 
    for satellite
    galaxies (left panel, blue), central galaxies (middle panel, red),
    and the total population (right panel). Contours and
    medians in bins of $\Delta$SFMS are shown for each population,
    with 25th/75th percentile error bars; the unity line is shown with
    dashes. Large median points are
    filled if more than 
    half are detected in \hi. \added{Satellite galaxies in the TZ
      are offset to lower median \dg \,  than central galaxies; a KS
      test on these two distributions produces $p=0.012$ implying they
      are significantly different from each other.}
    \label{fig:tzsat}}
\end{figure*}

While all galaxies in the TZ could naively be considered
``quenched'' based on their sSFR alone, we are interested in
separating the TZ into meaningful populations that may evolve on
different pathways. We begin by separating TZ galaxies by their
environmental identity as 
satellite or central galaxies, since we expect these populations to
evolve through different mechanisms. We use the SDSS DR7 group
catalog of \citet{yang07} to divide our sample into
satellite members of groups, central members of groups, and isolated
central galaxies, and we first consider the satellite galaxy population.

\subsection{Satellite galaxies in the TZ}
\label{sec:sat}

There are 160 satellite galaxies from xGASS in the TZ, of which 67 have
\hi \, detections. The left panel of Figure~\ref{fig:tzsat} shows the
\dg \, 
vs.~$\Delta$SFMS plot for all satellites, with medians
plotted in bins of $\Delta$SFMS. Compared with
central galaxies (in groups and in isolation), satellites have  median
\dg \, 
values which are smaller both on and below the SFMS
(above the SFMS, the central and satellite populations are
indistinguishable). The median offset in \dg \, 
between the 
\hi-detected TZ satellite and central galaxies is $\sim$0.47~dex, 
which is larger than the analogous offset in the SFMS
($\sim$0.10~dex).
\added{Quantitatively, the distributions of \dg \, for \hi-detected
  central and satellite galaxies in the TZ are different at the 98.8\%
  level ($p=0.012$) using a two-tailed $p$-value KS test. }
This reduced gas content of satellite galaxies
is broadly consistent with environmental stripping/quenching
scenarios,
where a reduction of gas content precedes a decrease in star formation
activity.

\begin{figure*}
  \centering
  \includegraphics[width=0.99\textwidth]{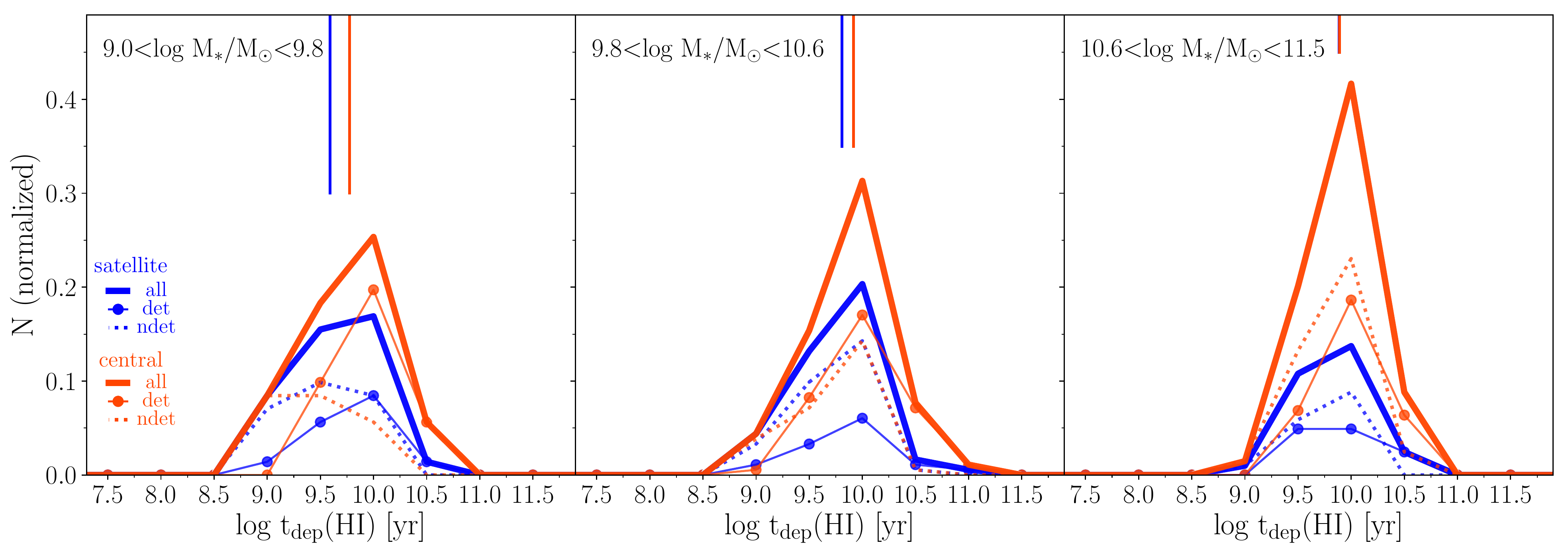}
  \caption{Normalized histograms of \hi \, depletion time for TZ galaxies
    which are centrals (red) and satellites (blue), divided into three
    bins of stellar mass.
    Thick lines show histograms of 
    all of the (\hi) detected and non-detected galaxies, thin lines
    show only detections, and dotted lines show only non-detections.
    Vertical lines are drawn at the median depletion time for each
    population (including detections and non-detections).
    \added{In the lowest mass bin the satellites (N=12) have shorter
      depletion times than centrals (N=25)and a KS test between these two
      distributions gives $p=0.26$ which is a weak difference.  The
      middle and high mass bins show less significant differences in
      their \hi \, depletion times.}
    \label{fig:tzcent}}
\end{figure*}

\added{While the \dg \, distributions of central
  and satellite galaxies in the TZ show statistically significant
  differences, it 
  is difficult to physically interpret any differences in \dg \, alone
  since it implicitly includes a dependence on \Mst \, (present in the
  \hi \, gas fraction shown in Figure~\ref{fig:sfgf}). In order to
  quantify this apparent quenching of 
satellite galaxies in a more physical sense, we instead consider the
distributions of their \hi \, depletion times
(using only galaxies with \hi \, detections). In 
the TZ, a KS test of the distributions of \hi \, depletion times for
central and satellite galaxies produces a two-sided p-value of
$p=0.13$, implying these samples are only weakly distinct from each
other.}

\added{
To dig deeper into the significant trends identified in \dg,
Figure~\ref{fig:tzcent} shows the \hi \, depletion times for TZ
galaxies now separated into three bins of stellar mass.}
In all bins, the distributions of \hi \, depletion times for
satellite and central TZ galaxies have similar widths of
$\sim$0.4~dex.  
In the lowest mass bin (with the highest \hi \,
detection fraction), satellite TZ galaxies have median \hi \,
depletion times of 10$^{9.6}$~yr, while central TZ galaxies have a
median of 10$^{9.8}$~yr. \added{This difference is not very large and
 a KS test on these two distributions of depletion times yields only
 $p=0.26$. The same galaxies have \dg \, distributions which are
 different by $p=0.07$, which is somewhat more significant.
The middle and high mass bins show even smaller differences between TZ
satellites and centrals, so this trend appears to be driven by the
galaxies at lower mass.
}
The reason why \hi \, depletion times show less of a difference than
\dg \, is most likely because changes in SFR and \hi \, content cancel
each other and/or increase the scatter. As shown in
Figure~\ref{fig:tzsat}, central and satellite galaxies are different
at nearly fixed $\Delta$SFMS, but their difference relative to the
diagonal dashed line (i.e., constant depletion time) is inevitably
less significant.

\added{While this difference in \hi \, depletion times is small
  ($p=0.012$), a systematically} lower HI content 
in satellites at fixed star formation rate (i.e., shorter depletion
times) than centrals is in line with a simple evolutionary picture
where 
satellite quenching is driven primarily by gas stripping. In this
scenario, satellite galaxies in the TZ (especially those at low mass)
have had their gas reservoirs reduced more rapidly and their star
formation is still declining to match. This points towards active
stripping of gas being more important in the evolution of satellites
than central galaxies, though the large scatter may indicate multiple
evolutionary paths also for the satellite population.

\subsection{Central galaxies in the TZ}
\label{sec:cent}

There are 297 central galaxies from xGASS in the TZ, of which 158 have
\hi \, detections. 
Galaxies in the TZ span a similarly wide range of \hi \, depletion
times compared with those in the SFMS or above it. As discussed in
Section~\ref{sec:sat}, the smaller gas reservoirs of satellite
galaxies in the TZ is consistent with a picture of environmental
effects driving them into the TZ.
In this section we now 
focus on the central galaxies, which we expect to be less affected by their
environment. Here we combine all central galaxies (those in
groups/clusters and those in isolation) into a single sample for
better statistics. This yields a sample of 297 central galaxies in the
TZ. We also carry out the following analysis on subsets of isolated
only and group central galaxies only, and find no significant
differences from the results discussed below.

\begin{figure}
  \centering
  \includegraphics[width=0.99\columnwidth]{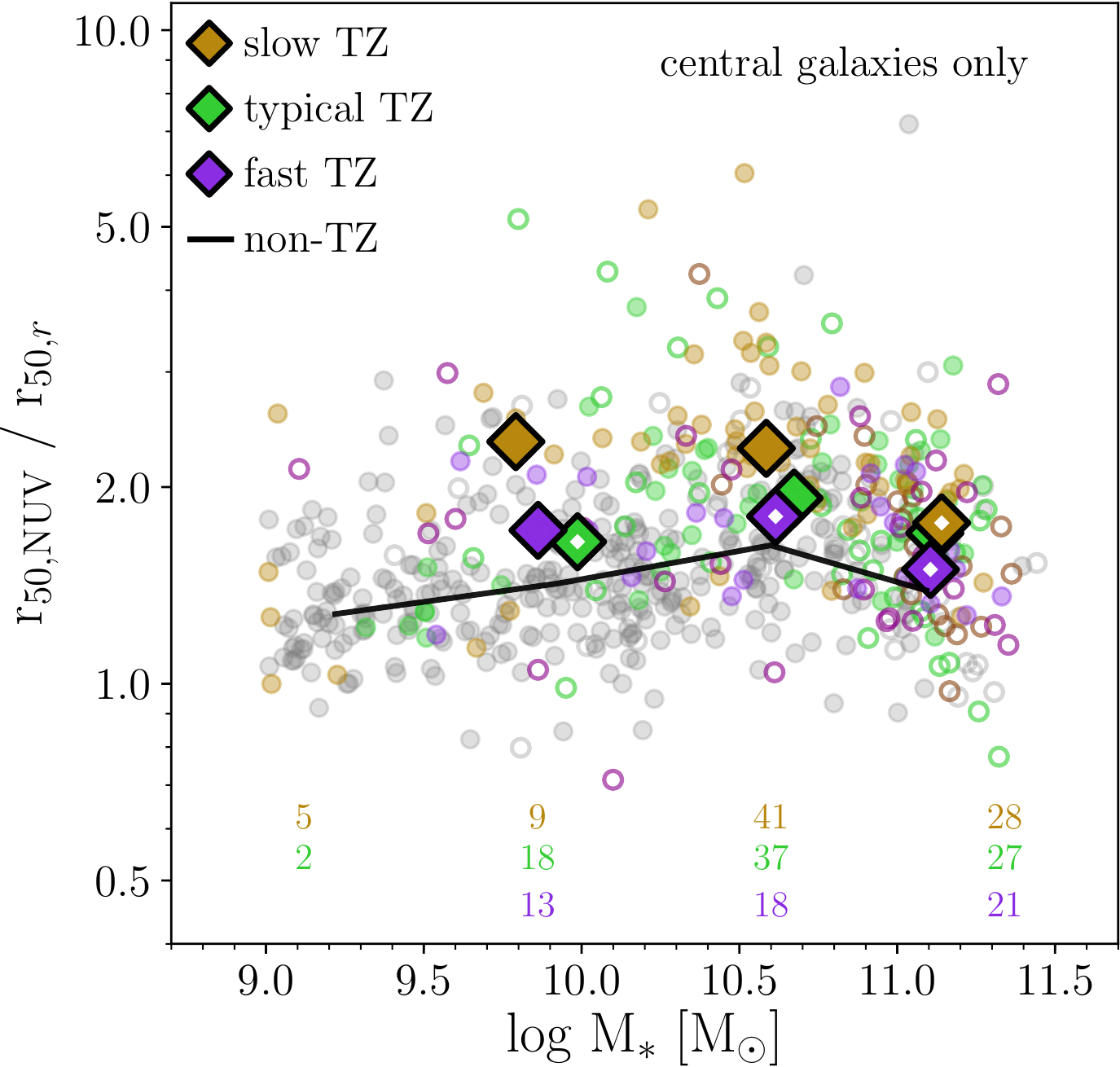} 
  \caption{Only central galaxies are plotted,
    using different colors for those in
    the TZ with short \hi \, depletion times (fast, purple), average
    depletion times (typical, green), and long (slow, brown). Grey
    symbols indicate central galaxies outside of the TZ. Solid symbols
    denote
    \hi \, detections, and open show upper limits of non-detections. Large
    diamonds show medians (including detections and non-detections) in
    bins of stellar mass (when fewer than 50\% are detected, the
    diamonds are open). Numbers at the bottom indicate how many
    galaxies are in each binned median.
    The black line connects medians of the non-TZ central galaxies.
    TZ central galaxies show a trend with the UV-to-optical size ratio
    such that those with longer (slower) depletion times have larger
    ratios.
    \label{fig:uvsize}}
\end{figure}

To explore this population, we divide the central galaxies in the TZ
into three roughly equal-sized populations based on their (\hi)
depletion times. ``Typical'' centrals have t$_\textrm{dep}$ between
$\sim$5 and $\sim$11~Gyr, and represent the middle trintile of the
distribution. Those with longer depletion times are considered
``slow,''  and shorter are considered ``fast.''  We also note that
this population of central TZ galaxies with ``typical'' \hi \,
depletion times is found along the 1:1 line shown in
Figure~\ref{fig:dd}, the ``fast'' are found above the unity line, and
the ``slow'' below.

We test for possible trends with a few fundamental observed
properties of the TZ 
centrals in these ``fast,'' ``typical,'' and ``slow''
populations. With regard to their environment, we consider the
density of nearby neighbors and their halo mass as given by abundance
matching in the group catalog of \citet{yang07}. When using the
density of galaxies per projected Mpc$^2$ within the 7th nearest 
neighbor, we find no significant difference between these three
populations of TZ centrals. Similarly, no trends are found with the
halo masses. Both of these results suggest that the gas content in
TZ centrals is not strongly affected by their environment, and that
this population may evolve through a variety of channels.

Looking internally, we also look for trends with average stellar
surface density, $\mu_*$, defined as \Mst/(2$\pi$R$_{50,z}^2$), where
the $z$ filter half-light radius $R_{50,z}$ is measured in kpc. All
three  populations of TZ centrals have larger average values than
non-TZ central galaxies, which is consistent with the typical
morphologies seen below the SFMS \citep[e.g.,][]{salim07}. However, we
find  no significant trends as a function of $\mu_*$. This further
shows the diversity and scatter within this population.

The only meaningful trend we identify appears when considering the
ratio of UV and optical 
size of the TZ centrals. We use the ratio of the half-light radius
measured in NUV to the half-light radius in the optical $r$
filter,  including only $\sim$75\% of galaxies that have been
resolved in the \textit{GALEX} images (i.e., $r$$>$5$''$). Here the
three TZ central populations show a consistent trend:  
those with the longest \hi \, depletion times also have the highest
average UV-to-optical size ratios, and the smallest \hi \, depletion
times have the smallest UV-to-optical size ratios.

\newcommand{\www}{0.115\textwidth}
\newcommand{\wwww}{0.105\textwidth}

\newcommand{\hz}{\hspace{39pt}}

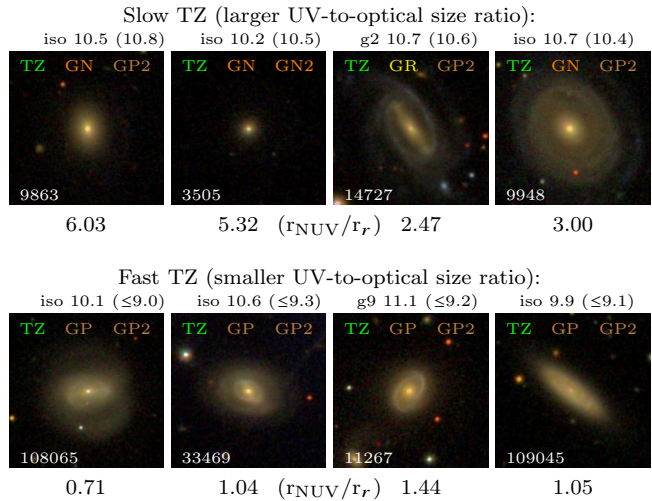
\begin{figure}
  \centering
  Slow TZ (larger UV-to-optical size ratio):\\
\FramedBoxx{\wwww}{iso~10.5~(10.8)\\ \textcolor{white}{.}}
\FramedBoxx{\wwww}{iso~10.2~(10.5)\\ \textcolor{white}{.}}
\FramedBoxx{\wwww}{g2~10.7~(10.6)\\ \textcolor{white}{.}}
\FramedBoxx{\wwww}{iso~10.7~(10.4)\\ \textcolor{white}{.}}
\begin{overpic}[width=\www]{9863_sdss_small2.jpg}
\put(7,87){\scriptsize\textcolor{green}{TZ}}
\put(36,87){\scriptsize\textcolor{orange}{GN}}
\put(67,87){\scriptsize\textcolor{brown}{GP2}}
\put(7,3){\scriptsize\textcolor{white}{9863}}
\end{overpic}
\begin{overpic}[width=\www]{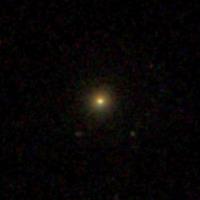}
\put(7,87){\scriptsize\textcolor{green}{TZ}}
\put(36,87){\scriptsize\textcolor{orange}{GN}}
\put(67,87){\scriptsize\textcolor{orange}{GN2}}
\put(7,3){\scriptsize\textcolor{white}{3505}}
\end{overpic}
\begin{overpic}[width=\www]{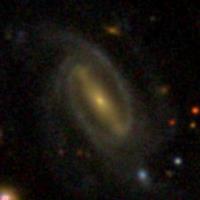}
\put(7,87){\scriptsize\textcolor{green}{TZ}}
\put(36,87){\scriptsize\textcolor{yellow}{GR}}
\put(67,87){\scriptsize\textcolor{brown}{GP2}}
\put(7,3){\scriptsize\textcolor{white}{14727}}
\end{overpic}
\begin{overpic}[width=\www]{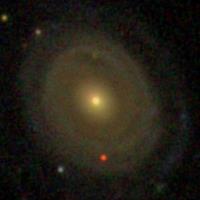}
\put(7,87){\scriptsize\textcolor{green}{TZ}}
\put(36,87){\scriptsize\textcolor{orange}{GN}}
\put(67,87){\scriptsize\textcolor{brown}{GP2}}
\put(7,3){\scriptsize\textcolor{white}{9948}}
\end{overpic}\\
  6.03 \hz  5.32 \, (r$_\textrm{NUV}$/r$_r$) \, 2.47 \hz  3.00 \\
  \vspace{10pt}
  Fast TZ (smaller UV-to-optical size ratio):\\
\FramedBoxx{\wwww}{iso~10.1~($\le$9.0)\\ \textcolor{white}{.}}
\FramedBoxx{\wwww}{iso~10.6~($\le$9.3)\\ \textcolor{white}{.}}
\FramedBoxx{\wwww}{g9~11.1~($\le$9.2)\\ \textcolor{white}{.}}
\FramedBoxx{\wwww}{iso~9.9~($\le$9.1)\\ \textcolor{white}{.}}
\begin{overpic}[width=\www]{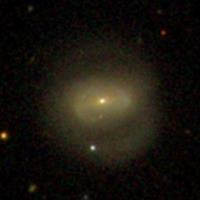}
\put(7,87){\scriptsize\textcolor{green}{TZ}}
\put(36,87){\scriptsize\textcolor{brown}{GP}}
\put(67,87){\scriptsize\textcolor{brown}{GP2}}
\put(7,3){\scriptsize\textcolor{white}{108065}}
\end{overpic}
\begin{overpic}[width=\www]{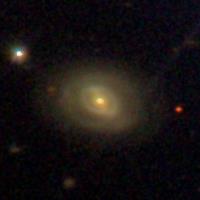}
\put(7,87){\scriptsize\textcolor{green}{TZ}}
\put(36,87){\scriptsize\textcolor{brown}{GP}}
\put(67,87){\scriptsize\textcolor{brown}{GP2}}
\put(7,3){\scriptsize\textcolor{white}{33469}}
\end{overpic}
\begin{overpic}[width=\www]{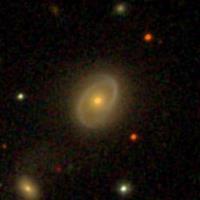}
\put(7,87){\scriptsize\textcolor{green}{TZ}}
\put(36,87){\scriptsize\textcolor{brown}{GP}}
\put(67,87){\scriptsize\textcolor{brown}{GP2}}
\put(7,3){\scriptsize\textcolor{white}{11267}}
\end{overpic}
\begin{overpic}[width=\www]{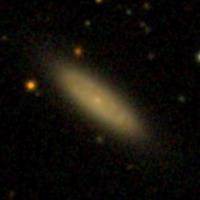}
\put(7,87){\scriptsize\textcolor{green}{TZ}}
\put(36,87){\scriptsize\textcolor{brown}{GP}}
\put(67,87){\scriptsize\textcolor{brown}{GP2}}
\put(7,3){\scriptsize\textcolor{white}{109045}}
\end{overpic}\\
  0.71 \hz 1.04 \, (r$_\textrm{NUV}$/r$_r$) \, 1.44 \hz  1.05\\
  \caption{SDSS color images of example central galaxies from the TZ,
    selected for their slow (top row) or fast (bottom row) \hi \,
    depletion times. TZ central galaxies with long depletion times
    typically show star formation activity in their outskirts and have
    a larger UV-to-optical size ratio, while those with short depletion
    times more often have central star formation and small ratios of
    UV-to-optical size. Galaxies are labeled as in Figure~\ref{fig:ex},
    with their environmental identity, stellar mass, and \hi \,
    depletion time.  The UV-to-optical size ratio
    (r$_\textrm{NUV}$/r$_r$) of each galaxy is listed underneath its thumbnail.
    \label{fig:tdepex}}
\end{figure}

Figure~\ref{fig:uvsize} shows this UV-to-optical size ratio vs.~stellar mass
for central galaxies in our sample. Those in the TZ are plotted in
colors corresponding to their depletion time; all non-TZ central
galaxies are shown as grey dots. There is a weak trend evident with
stellar mass such that more massive galaxies have larger UV-to-optical
size ratio. Additionally, most of the central
galaxies with the largest UV-to-optical size ratios are found in the TZ.
However, most intriguing is the trend amongst the TZ galaxies such
that those with longer depletion time (i.e., ``slow'') have larger
typical UV-to-optical size ratios. This is most clearly visible in the 
stellar mass bin centered around 10$^{10.5}$\msun, where there are
more galaxies of each population included in the median. For a visual
example, Figure~\ref{fig:tdepex} shows sample images of TZ galaxies in
the ``slow'' (with large UV-to-optical size ratio) and ``fast'' (with
small UV-to-optical size ratio) subsets. In particular, GASS~3505 was
identified 
by \citet{gereb16} as a prototypical example of an 
``\hi-excess'' galaxy -- a population  which has large \hi \,
reservoirs but insignificant amounts of star formation
\citep{gereb18}.

This correlation between depletion time and UV-to-optical size ratio
suggests that 
the star formation in the ``slow'' TZ galaxies is more radially
extended than in typical galaxies. It may mean that the excess \hi \,
reservoir is also more spatially extended and inaccessible to star
formation. 
 An inaccessible \hi \, reservoir may lead to low star formation
efficiency; $\sim$90\% of the ``slow'' TZ central galaxies have depletion
times longer than a Hubble time (and $\sim$15\% are more than twice as
long as a Hubble time). 
Intriguingly, for the $\sim$40\% of TZ galaxies with molecular
observations there is no correlation between the
molecular and atomic gas depletion times, and all three populations
have average molecular gas depletion times of $\sim$3~Gyr. 
The apparent tightness of the relation between
$\Delta$SFMS and \dgg \, (shown in the right panel of
Figure~\ref{fig:dd}) supports the fact that SFR proceeds whenever
molecular gas is available \citep[see, e.g., ][]{schruba11}; the
inaccessible \hi \, may be in an extended 
low-density configuration which prevents it from converting to molecular
gas and forming stars. 
These ``slow'' galaxies may still be forming
stars (albeit slowly) long after the galaxies in the SFMS have
exhausted their gas, fully
quenched, and joined the RS.

\section{Discussion}
\label{sec:discussion}

Galaxies in the TZ below the SFMS do not appear to make up a unified
population, as many external and internal mechanisms can drive
galaxies into (or out of) this so-called ``green valley.'' Compared to
SFMS galaxies, TZ galaxies 
have a similar range of \hi \, gas fractions and longer depletion
times, and are found across all environments, providing a large range
of available evolutionary pathways.

When considering the role of external environmental effects, we have
demonstrated that satellite galaxies with low \dg \, (see
Section~\ref{sec:sat}) may be examples of some TZ galaxies which
have experienced a cessation of star formation following the ram
pressure stripping of their gas by a cluster
environment. \citet{vulcani15} also find a weak connection between
environment and quenched galaxies, but also identify a population of
galaxies with intermediate optical colors which are not rapidly
quenched but instead are experiencing declining SFH over long time
scales ($>$1~Gyr). 

Our sample of TZ galaxies have average \hi \, depletion times of
$\sim$10~Gyr (see Section~\ref{sec:tdep}), suggesting that
this population is not rapidly moving toward the RS through
secular evolution alone. This population of long depletion time TZ
galaxies is in contrast with the expectations from other studies of
the stellar populations of TZ galaxies. For example, while
\citet{smethurst15} find evidence for multiple evolutionary paths
through the TZ, their scenario only requires 1$-$1.5~Gyr for a galaxy to
transit the TZ. Similarly, using the EAGLE hydrodynamic simulation,
\citet{trayford16} selected TZ galaxies based on their $u$-$r$ colors
and found that they spend less than 2~Gyr transiting between the SFMS
and RS. This apparently rapid quenching and disagreement with our
findings is largely a result of selecting TZ galaxies (and determining
their SFRs) based only on optical colors. As effectively demonstrated
by Figure~4 in \citet{salim14}, TZ and RS galaxies have
indistinguishable optical colors and it is difficult to make
meaningful distinctions between these populations without UV
observations.

Another formation scenario for TZ galaxies is that they have depleted
their gas reservoirs and are fading onto the RS. The results of
\citet{schawinski14} support the idea of multiple evolutionary
pathways through the TZ, but argue that the star formation rates are
responding to changes in their (un-observed) gas supplies. They claim
that late-type TZ galaxies have had their cosmic gas supply cut off,
and that early-type TZ galaxies have recently experienced a rapid
quenching as a result of the destruction of their gas supply by a
possible major merger. (We found no dependence of any of our
relationships on $\mu_*$, a proxy for morphology, as discussed in
Section~\ref{sec:cent}). 
 \citet{bremer18} also argue that an insufficient
gas supply is responsible for galaxies rapidly ($\sim$1$-$2~Gyr)
evolving through the TZ. While our work does confirm that TZ
galaxies have lower average \hi \, gas fractions than those in the
SFMS, 
the range of gas fractions is actually larger in the TZ, and their
long gas depletion times ($\sim$10~Gyr on average) are not consistent
with rapid quenching scenarios.

When selecting TZ galaxies based on  UV and optical observations,
others have also found TZ galaxies still 
forming stars at a sustainable rate. In particular, \citet{fang12}
identified a population of ``extended star-forming early-type
galaxies'' (ESF-ETGs) which were selected using their global
photometry on a UV-optical color-magnitude diagram. While consistent
with TZ in a global sense, these ESF-ETGs look red and morphologically
early-type at their centers but their outskirts are blue and show UV
emission from young stars. \citet{fang12} suggest that this population
of TZ galaxies may persist for several Gyrs before fading into the RS,
consistent with our results.

Studies of the gas content in star-forming galaxies at higher
redshifts have identified similar trends to those discussed in this
work. Notably, the Plateau de Bure high-z Blue Sequence Survey
\citep[PHIBSS,][]{genzel15,tacconi18} observed molecular gas
masses in 
32 galaxies at $z=1.2$ and $2.2$ with stellar masses above
$10^{10.4}$\msun, and found
that their molecular gas fraction generally decreases with time and
that their molecular gas depletion times increase further below the
SFMS, independent of stellar mass. 
More recently, the low redshift (z=0.5-0.8) sample of PHIBSS2 
galaxies between 10$^{10}$$<$\Mst$/$\msun$<$10$^{11.4}$ showed that 
this trend for molecular gas depletion time to increase below
the SFMS also continues towards lower redshifts 
\citep{freundlich19}. These 
relations are also consistent with the results of
\citet{saintonge12,saintonge17}, who used the COLD~GASS and xCOLD~GASS
samples (which are also used in this work and consist of direct CO
observations rather than indirect gas indicators used at higher
redshifts) to determine that the 
molecular gas depletion time increases by $\sim$0.3~dex between
galaxies in the SFMS and the TZ. Our findings that local galaxies
below the SFMS have longer \hi \, depletion times fits naturally into
this picture, and we have explored a few of the features and 
mechanisms at play in this evolution.

%
%

Many evolutionary paths lead through the TZ, and
today's TZ galaxies appear to be able to evolve in almost any
direction in Figure~\ref{fig:dd}. As a population, our TZ galaxies
do not appear to represent an intermediate stage between the SFMS and 
the RS, but rather are a diverse population, as seen most clearly
through observations of their cold gas content.

\section{Summary}
\label{sec:summary}

In this work we have used the xGASS and xCOLD~GASS samples to explore
the cold gas 
properties of galaxies in the TZ below the SFMS. Our main findings are:
\begin{itemize}
\item Galaxies in the SFMS have large scatter in \hi \, gas fraction
  scaling relation, but tighter in \H2. There is no corresponding
  ``\hi \, main sequence'' of galaxies in the SFMS.
\item Compared with galaxies in the SFMS, the \hi \, depletion times of
  galaxies in the TZ are \textbf{longer}, suggesting it is not a
  uniformly rapidly quenching/evolving population, and evolves more
  slowly than the SFMS.
\item \added{Satellite galaxies in the TZ show a trend to be more
  gas poor than TZ centrals, especially in terms of \dg. This
  trend is weaker when measured in \hi \, depletion time and appears to
  originate from the lowest mass galaxies (\Mst/\msun$<$$10^{9.8}$),}
  which is generally consistent with environmental stripping
  processes removing the gas.
\item Central galaxies in the TZ show a trend between their \hi \,
  depletion time and 
  UV-to-optical size ratios, suggesting that they have remained gas
  rich as a result of inefficient star formation proceeding slowly in
  their outer disks. Visual inspection of this population shows a
  variety of disk galaxies with strong bulges and  blue outer disks.
\end{itemize}

In order to further improve our understanding of the evolutionary
pathways available to galaxies in the TZ, we need enhanced spatial and
temporal resolution of their star formation activity. Given the
spatial distributions of UV emission discussed in
Section~\ref{sec:cent}, followup imaging in \Ha or UV would allow us
to better understand if star formation in TZ galaxies is proceeding
in different ways than in the SFMS. Re-constructing the star formation
histories of TZ galaxies would be invaluable in determining whether
they have been a long-lived TZ population or have recently experienced
a decline in SFR.

Better separating the mass contributions from the disk (star-forming,
associated with gas reservoir) 
and the bulge (passive) in the xGASS galaxies would also provide a
clearer picture of the evolutionary context of the TZ \citep[e.g.,][]{cook19}. Bulge growth
can contribute to a galaxy's decline from the SFMS to the TZ, and in
cases where the depletion times of the bulge and disk can be mapped
separately \citep[e.g.,][]{lin17}, it is clear that bulge quenching
and disk quenching are important and somewhat independent processes
which drive galaxies away from the SFMS. A better separation of bulge
and disk masses will help clarify these evolutionary pathways.

\section*{Acknowledgements}


We thank Toby Brown, Katinka Ger\'eb, and Robin Cook for helpful
discussions, \added{and the anonymous referee for their careful
  reading and thoughtful suggestions which
  have helped improve this work.}


SJ, BC, and LC acknowledge support from the Australian Research
Council's Discovery Project funding scheme (DP150101734).
Parts of this research were conducted by the Australian Research
Council Centre of Excellence for All Sky Astrophysics in 3 Dimensions
(ASTRO 3D), through project number CE170100013.
LC is the recipient of an Australian Research Council Future
Fellowship (FT180100066) funded by the Australian Government.

This research has made use of NASA's Astrophysics Data System, and
also the 
NASA/IPAC Extragalactic Database (NED), which is operated by the Jet
Propulsion Laboratory, California Institute of Technology, under
contract with the National Aeronautics and Space Administration. This
research has also made extensive use of the amazingly invaluable Tool for
OPerations on  Catalogues And Tables
\citep[TOPCAT\footnote{\url{http://www.starlink.ac.uk/topcat/}},][]{taylor05}.

Funding for SDSS-III has been provided by the Alfred P. Sloan
Foundation, the Participating Institutions, the National Science
Foundation, and the U.S. Department of Energy Office of Science. The
SDSS-III web site is \url{http://www.sdss3.org/}.

SDSS-III is managed by the Astrophysical Research Consortium for the
Participating Institutions of the SDSS-III Collaboration including the
University of Arizona, the Brazilian Participation Group, Brookhaven
National Laboratory, Carnegie Mellon University, University of
Florida, the French Participation Group, the German Participation
Group, Harvard University, the Instituto de Astrofisica de Canarias,
the Michigan State/Notre Dame/JINA Participation Group, Johns Hopkins
University, Lawrence Berkeley National Laboratory, Max Planck
Institute for Astrophysics, Max Planck Institute for Extraterrestrial
Physics, New Mexico State University, New York University, Ohio State
University, Pennsylvania State University, University of Portsmouth,
Princeton University, the Spanish Participation Group, University of
Tokyo, University of Utah, Vanderbilt University, University of
Virginia, University of Washington, and Yale University.











\bsp	
\label{lastpage}
\end{document}